\documentclass[11pt]{article}
\usepackage[T1]{fontenc}
\usepackage{textcomp}
\usepackage{lmodern}
\usepackage{amsmath,amssymb,amsthm,mathtools}
\usepackage[round,authoryear]{natbib}
\usepackage{booktabs,threeparttable,tabularx,array,graphicx}
\usepackage[ruled,vlined,linesnumbered,algo2e]{algorithm2e}
\usepackage{titlesec,needspace,placeins,xurl}
\usepackage[letterpaper,margin=1in]{geometry}
\usepackage[colorlinks=true,linkcolor=blue,citecolor=blue,urlcolor=blue]{hyperref}
\providecommand{\doi}[1]{doi: \href{https://doi.org/#1}{\nolinkurl{#1}}}
\hypersetup{pdftitle={Sharp identification and finite-sample inference for AUC from binned data},pdfauthor={Hengyao Xu, Tao Tan}}
\titleformat{\section}{\normalfont\large\bfseries}{\thesection}{0.65em}{}
\titleformat{\subsection}{\normalfont\normalsize\bfseries}{\thesubsection}{0.65em}{}

\newtheorem{lemma}{Lemma}
\newtheorem{theorem}{Theorem}
\newtheorem{proposition}{Proposition}
\newtheorem{corollary}{Corollary}
\theoremstyle{remark}
\newtheorem{remark}{Remark}
\theoremstyle{plain}

\newcommand{\E}{\mathbb{E}}
\newcommand{\Prob}{\mathbb{P}}
\newcommand{\Var}{\operatorname{Var}}
\newcommand{\ind}{\mathbf{1}}
\newcommand{\clip}{\operatorname{clip}}
\newcommand{\act}{\mathcal{O}}
\newcommand{\sat}{r_{\mathrm{sat}}}
\newcommand{\DKW}{\mathrm{DKW}}
\newcommand{\KL}{\mathrm{KL}}
\newcommand{\placement}[2]{}
\newif\ifreviewmode\reviewmodetrue

\title{\bfseries Sharp identification and finite-sample inference for AUC from binned data\textsuperscript{\textdagger}}
\author{Hengyao Xu, Tao Tan\textsuperscript{*}\\
College of Mathematics and Systems Science, Xinjiang University,\\Urumqi, Xinjiang, China}
\date{}
\makeatletter
\renewcommand{\@makefnmark}{}
\renewcommand{\@makefntext}[1]{\parindent 0pt\noindent#1}
\makeatother
\def\arxivversion{}

\begin{document}
\maketitle
\footnotetext{\textsuperscript{*}Corresponding author. Email: \texttt{ttan303@163.com}.}
\footnotetext{\textsuperscript{\textdagger}This research was supported by the Natural Science Foundation of Xinjiang Uygur Autonomous Region (No. 2025D14015) and the National Natural Science Foundation of China (No. 72571102).}
\begin{abstract}
Ordered bins preserve rankings across bins but hide rankings within them. We study what these bins reveal about the area under the receiver operating characteristic curve (AUC). A shared Brier information budget limits how well scores can distinguish groups within bins. We derive a sharp AUC interval under this budget. Every value in the interval can occur. The optimal allocation yields the endpoints and the minimum budget at which the lower endpoint meets or falls below a benchmark. For two independent samples, we construct confidence intervals for the identified set at a fixed budget and simultaneous bands across budgets. A certified test controls finite-sample size at a prespecified budget and threshold. Simulated power more than doubled over paired Kullback--Leibler bounds in two designs and remained low under severe imbalance. In PTB-XL, the sample lower endpoint first reached the working AUC threshold of 0.70 at a budget of 0.1483.
\end{abstract}
\medskip

\section{Introduction}\label{sec:intro}

In the PTB-XL electrocardiography database, we asked how well age distinguishes patients with reports labeled myocardial infarction (MI) from those labeled normal electrocardiogram (NORM) \citep{Wagner2020,PTBXLv103}. Across six ordered age bins, the sample mid-rank area under the receiver operating characteristic (ROC) curve (AUC) was $0.7420$. The sharp sample interval without a within-bin information restriction was $[0.6657,0.8183]$. It straddled the working threshold of $0.70$. Bins order observations across categories but hide comparisons within categories. How much hidden information would make an AUC of $0.70$ or less compatible with these data?

A tie receives half credit in the standard two-sample rank probability \citep{FayProschan2010}, but binning hides the score comparisons behind such ties. Bounds from finite ROC operating points constrain area between observed thresholds \citep{VanDenHout2003,ZhangMueller2005,MuellerZhang2006}. ROC methods for measurements below detection limits address a related loss of ordering in a censored tail \citep{Bandos2017,Bandos2017supp}. These methods bound performance from threshold or category summaries without indexing the bounds by within-bin information.

\ifdefined\arxivversion\newpage\fi
Recent work addresses related questions. \citet{Cormode2023} derive accuracy bounds for federated classifier evaluation, including ROC-AUC under privacy guarantees. \mbox{\citet{XuCormode2025}} approximate ROC and precision--recall curves from score quantiles using secure aggregation and distributed differential privacy. \citet{Schalk2024} estimate ROC curves, AUCs, and confidence intervals across sites by combining a distributed ROC-GLM with differentially private survivor-function approximations. Their simulations show that privacy noise can affect accuracy. They estimate performance from distributed score data. We instead hold the ordered bin probabilities fixed and characterize the full set of compatible AUCs. That set-valued target is a partial-identification problem \citep{Manski2003}. \citet{DalyGrafstein2026} use clinically relevant restrictions in a nonparametric Bayesian model to narrow the ATE region for nonignorable missing binary outcomes. These include instrumental-variable validity and assumptions about the direction of missingness bias. We restrict within-bin score discrimination.

We measure within-bin information by the proportional reduction in equal-prior Brier risk for predicting group membership when the score is observed rather than only its bin \citep{Reid2011}. \citet{GongChan2026} develop multiple sensitivity models for generalized estimating equations with multiple missing variables. Here a shared sensitivity budget caps the weighted average of per-bin Brier risk reductions, with positive weights fixed in advance. Varying this cap shows how AUC conclusions depend on hidden discrimination.

Classical one-bin extremal results give the least Brier information needed for a specified departure from a within-bin win probability of one half \citep{Cornilly2018,Pham2011}. We use these costs to characterize the sharp AUC interval under a shared budget, and every value between its endpoints is attainable. The optimal allocation equates marginal AUC gains per unit budget across overlapping bins, yielding a low-budget closed form and an explicit saturation budget. The critical budget is the smallest allowance at which the lower endpoint reaches or falls below a benchmark, if one exists \citep{Imbens2004,Kline2013,Masten2020}.

The bin probabilities must also be estimated. Paired Kullback--Leibler bounds give a confidence interval for the population identified set at a fixed budget, while projected distribution-function bounds give simultaneous bands across budgets. Inverting the lower bounds gives confidence limits for critical budgets. For multinomial functionals, \citet{Sachs2025} derive exact small-sample p-values and a Monte Carlo approximation. At a prespecified budget and threshold, our test uses a fixed witness to bound the sample lower endpoint and certifies the tail bound over null configurations. Under \hyperlink{assumption:M1}{(M1)}--\hyperlink{assumption:M4}{(M4)} and the stated numerical enclosure conditions, its finite-sample level is at most $0.025$.

In PTB-XL, the sample lower endpoint first reached $0.70$ at a budget of $0.1483$. Sections 2--4 establish identification, optimal allocation, and inference. Sections 5--7 report simulations and three applications, then discuss interpretation and limitations.

\section{Model and Sharp Identification}\label{sec:identification}

\subsection{Binned data and the target AUC}\label{subsec:model}

Let $I_1,\ldots,I_J$ ($J\ge 1$ fixed and finite) be a prespecified, strictly ordered partition of the score support on $\mathbb{R}$ into intervals, each containing a nonempty open interval, with a fixed endpoint convention and bin-index map $B:\mathbb{R}\to\{1,\ldots,J\}$.

\begin{list}{}{\setlength{\labelwidth}{2.2em}\setlength{\labelsep}{0.5em}\setlength{\leftmargin}{2.7em}\setlength{\itemindent}{0pt}\setlength{\topsep}{4pt}\setlength{\itemsep}{3pt}\setlength{\parsep}{0pt}\renewcommand{\makelabel}[1]{#1\hfil}}
\item[\textbf{\hypertarget{assumption:M1}{(M1)}}] \emph{Sampling and atomless scores.}
Two independent iid samples $X_1,\ldots,X_{n_0}\sim P_0$ and $Y_1,\ldots,Y_{n_1}\sim P_1$ ($n_0,n_1\in\mathbb{Z}_{\ge 1}$) are drawn from atomless score distributions $P_0,P_1$ on $\mathbb{R}$. Only the bin counts
\[
 N_{0j}=\sum_{s=1}^{n_0}\ind\{B(X_s)=j\},\qquad
 N_{1j}=\sum_{t=1}^{n_1}\ind\{B(Y_t)=j\},\qquad j=1,\ldots,J,
\]
are observed, yielding empirical proportions $\hat p_j=N_{0j}/n_0$ and $\hat q_j=N_{1j}/n_1$. Under~\hyperlink{assumption:M1}{(M1)}, $N_0\sim\mathrm{Multinomial}(n_0,\boldsymbol{p})$ and $N_1\sim\mathrm{Multinomial}(n_1,\boldsymbol{q})$ independently, where $p_j=P_0(I_j)$ and $q_j=P_1(I_j)$ lie in the simplex $\Delta_J=\{w\in[0,1]^J:\sum_{j=1}^J w_j=1\}$.

\item[\textbf{\hypertarget{assumption:M2}{(M2)}}] \emph{Target parameter and within-bin quantities.}
The target parameter is the continuous-score Mann--Whitney functional \citep{Hanley1982} $\theta=\Prob(X_1<Y_1)$. Define collision masses $c_j=p_j q_j$ and the active overlap set $\act(\boldsymbol{p},\boldsymbol{q})=\{j\in\{1,\ldots,J\}:c_j>0\}$ (abbreviated $\act$ at the true $(\boldsymbol{p},\boldsymbol{q})$). For $j\in\act$, let $P_{0j},P_{1j}$ be the conditional score distributions on $I_j$, $M_j=(P_{0j}+P_{1j})/2$ their equal-prior mixture, $\eta_j=dP_{1j}/d(P_{0j}+P_{1j})$ the equal-prior posterior probability of group~$1$, and
\[
 v_j=\Prob(X_1<Y_1\mid X_1\in I_j,\,Y_1\in I_j),\qquad
 b_j=4\Var_{M_j}(\eta_j)=\frac{1}{2}\int \frac{(f_{1j}-f_{0j})^2}{f_{0j}+f_{1j}}\,d\mu,
\]
where $f_{0j},f_{1j}$ are densities with respect to a common dominating measure $\mu$, taken with zero integrand when $f_{0j}+f_{1j}=0$.

\item[\textbf{\hypertarget{assumption:M3}{(M3)}}] \emph{Null-collision convention.}
For $j\notin\act$, set $b_j=0$ and $v_j=1/2$. When one group has zero mass in $I_j$, its formal conditional distribution is taken equal to the other group's conditional distribution; when both vanish, both take a common atomless law supported on an open subinterval of $I_j$.
\end{list}

A case in bin $j$ outranks a control in bin $i$ whenever $i<j$. Only the diagonal comparisons depend on $\boldsymbol{v}$, so the AUC functional is
\begin{equation}\label{eq:T_decomp}
 T(\boldsymbol{p}',\boldsymbol{q}';\boldsymbol{v})=\sum_{i<j}p'_iq'_j+\sum_jp'_jq'_jv_j,
 \qquad \mathbf{K}_{\boldsymbol{v}}(i,j)=\ind\{i<j\}+v_i\ind\{i=j\}.
\end{equation}
Using the collision terms in \hyperlink{assumption:M2}{(M2)}, \eqref{eq:T_decomp} gives $T(\boldsymbol{p}',\boldsymbol{q}';\boldsymbol{v})=\boldsymbol{p}'^{\mathsf T}\mathbf{K}_{\boldsymbol{v}}\boldsymbol{q}'$ and $\theta=T(\boldsymbol{p},\boldsymbol{q};\boldsymbol{v})$ at the true within-bin vector $\boldsymbol{v}$. Setting $\boldsymbol{v}=\mathbf{1}/2$ defines the mid-rank AUC $m(\boldsymbol{p}',\boldsymbol{q}')=T(\boldsymbol{p}',\boldsymbol{q}';\mathbf{1}/2)$, abbreviated $m$ at $(\boldsymbol{p},\boldsymbol{q})$. Letting each $v_j$ vary freely over $[0,1]$ yields the unrestricted sharp interval $[L_{\min},U_{\max}]=[m-R_{\max},m+R_{\max}]$, where $L_{\min}=\sum_{i<j}p_iq_j$ and $R_{\max}=\frac12\sum_j c_j$; its width $\sum_j c_j$ coincides with the unbudgeted within-bin bound in Lemma~3 of Section~5.1 of \citet{Cormode2023}. Throughout, population functionals omit hats and sample plug-ins at $(\hat{\boldsymbol{p}},\hat{\boldsymbol{q}})$ carry hats.

\subsection{Within-bin Brier information}\label{subsec:brier}

Inside an overlapping bin $j\in\act$, predicting class membership without access to the continuous score under equal priors incurs baseline Brier risk $1/4$, whereas Bayes-optimal prediction using $\eta_j$ attains risk $\E_{M_j}[\eta_j(1-\eta_j)]=1/4-\Var_{M_j}(\eta_j)$. Thus $b_j=4\Var_{M_j}(\eta_j)\in[0,1]$ equals the relative reduction in baseline equal-prior Brier risk achieved by observing the continuous score within $I_j$ \citep{Reid2011}. Zero information ($b_j=0$) holds if and only if $P_{0j}=P_{1j}$, which implies $v_j=1/2$, whereas symmetric conditional differences can leave $v_j=1/2$ even when $b_j>0$.

A shared budget limits the weighted average of these risk reductions across bins.

\begin{list}{}{\setlength{\labelwidth}{2.2em}\setlength{\labelsep}{0.5em}\setlength{\leftmargin}{2.7em}\setlength{\itemindent}{0pt}\setlength{\topsep}{4pt}\setlength{\itemsep}{3pt}\setlength{\parsep}{0pt}\renewcommand{\makelabel}[1]{#1\hfil}}
\item[\textbf{\hypertarget{assumption:M4}{(M4)}}] \emph{Prespecified shared budget.}
Fix strictly positive weights $\boldsymbol{a}=(a_1,\ldots,a_J)\in\Delta_J$ ($a_j>0$, $\sum_{j=1}^J a_j=1$) prior to observing data. Bins with $p_j=0$, $q_j=0$, $\hat p_j=0$, or $\hat q_j=0$ are never dropped and $\boldsymbol{a}$ is never renormalized. The population within-bin Brier information is $\rho=\sum_{j=1}^J a_j b_j\in[0,1]$, and the sensitivity model posits $\rho\le r$ for a budget $r\in[0,1]$.
\end{list}

For example, $r=0.04$ allows at most a four-percent weighted average reduction relative to the equal-prior Brier baseline. The weights set the relative costs of hidden discrimination across bins. The number of cross-group pairs affected by that discrimination is governed separately by the collision masses $c_j$. At zero budget, overlapping bins contribute their mid-rank terms. At $r=1$, within-bin orderings are unrestricted. The observed counts determine neither $b_j$ nor $\rho$.

To express the budget as a constraint on win probabilities, we use the classical single-bin extremum given in Theorem~2.2 and Section~3.2 of \citet{Cornilly2018}.

\begin{lemma}\label{lem:A}
In an overlapping bin $j\in\act$, put $d=|v_j-1/2|\in[0,1/2]$. The minimum standardized Brier information $b_j$ compatible with either signed displacement $v_j-1/2=\pm d$ is
\begin{equation}\label{eq:C_cost}
 C(d)=\begin{cases}
 3d^2,& 0\le d\le 1/3,\\[3pt]
 1-\dfrac{2}{3}\sqrt{3-6d},& 1/3<d\le 1/2.
 \end{cases}
\end{equation}
Every displacement in $[-1/2,1/2]$ attains this minimum under atomless conditional laws supported on any open subinterval of $I_j$. On $[0,1/2]$, $C$ is continuous, strictly increasing, and strictly convex, with $C(0)=0$, $C(1/3)=1/3$, $C'(1/3-)=C'(1/3+)=2$, and $C(1/2)=1$.
\end{lemma}

The quadratic branch also follows from the Gini mean difference bound of \citet{Pham2011}. The \hyperlink{proof:lem:A}{proof of Lemma~\ref*{lem:A}} in Appendix~\ref*{app:proofs} gives a self-contained derivation and constructs attaining conditional laws.

\subsection{Sharp identification under a shared budget}\label{subsec:sharp}

Lemma~\ref{lem:A} gives a necessary cost for each within-bin displacement. It also gives conditional laws that attain this cost, which can be chosen separately in each bin and combined with the prescribed bin probabilities. This construction leads to the feasible sets
\[
 V_r=\left\{\boldsymbol{v}\in[0,1]^J:\sum_{j=1}^J a_j C(|v_j-1/2|)\le r\right\},\qquad
 D_r=\left\{\boldsymbol{d}\in[0,1/2]^J:\sum_{j=1}^J a_j C(d_j)\le r\right\},
\]
and, for any $\boldsymbol{p}',\boldsymbol{q}'\in\Delta_J$, the maximal radius and interval endpoint functionals
\begin{equation}\label{eq:R_L_U_def}
\begin{aligned}
 R(\boldsymbol{p}',\boldsymbol{q}',\boldsymbol{a};r)&=\max_{\boldsymbol{d}\in D_r}\sum_{j=1}^J p'_j q'_j d_j,\\
 L(\boldsymbol{p}',\boldsymbol{q}',\boldsymbol{a};r)&=m(\boldsymbol{p}',\boldsymbol{q}')-R(\boldsymbol{p}',\boldsymbol{q}',\boldsymbol{a};r),\\
 U(\boldsymbol{p}',\boldsymbol{q}',\boldsymbol{a};r)&=m(\boldsymbol{p}',\boldsymbol{q}')+R(\boldsymbol{p}',\boldsymbol{q}',\boldsymbol{a};r).
\end{aligned}
\end{equation}
At the population marginals $(\boldsymbol{p},\boldsymbol{q})$, write $R(r)=R(\boldsymbol{p},\boldsymbol{q},\boldsymbol{a};r)$, $L(r)=L(\boldsymbol{p},\boldsymbol{q},\boldsymbol{a};r)$, and $U(r)=U(\boldsymbol{p},\boldsymbol{q},\boldsymbol{a};r)$.

\begin{theorem}\label{thm:B}
Under \hyperlink{assumption:M1}{(M1)}--\hyperlink{assumption:M4}{(M4)}, the sharp identified set of continuous-score AUC values compatible with bin marginals $(\boldsymbol{p},\boldsymbol{q})$ and shared Brier information $\rho\le r$ is the closed interval
\[
 T(\boldsymbol{p},\boldsymbol{q};V_r)=[L(r),U(r)]=[m-R(r),\,m+R(r)],
\]
and every $\theta\in[L(r),U(r)]$ is attained by atomless score laws $(P_0,P_1)$ inducing exact bin probabilities $(\boldsymbol{p},\boldsymbol{q})$ and satisfying $\rho\le r$.
\end{theorem}

The unrestricted bound of \citet{Cormode2023} allows arbitrary within-bin ordering; Theorem~\ref{thm:B} identifies the values compatible with a shared information budget.

For any feasible score laws, Lemma~\ref{lem:A} places their win vector in $V_r$. To attain any value between the endpoints, take a maximizer $\boldsymbol{d}^*\in D_r$ with $d_j^*=0$ for $j\notin\act$ and set $v_j(t)=1/2+t d_j^*$ for $t\in[-1,1]$. By \eqref{eq:C_cost}, the cost in each bin is at most $C(d_j^*)$. Assign conditional laws achieving these wins to their bins and combine them with masses $\boldsymbol{p}$ and $\boldsymbol{q}$ in the two groups. Using the endpoints in \eqref{eq:R_L_U_def}, the resulting AUC runs from $L(r)$ to $U(r)$ as $t$ varies. The \hyperlink{proof:thm:B}{proof of Theorem~\ref*{thm:B}} in Appendix~\ref*{app:proofs} establishes sharpness and attainability.

\section{Optimal allocation and sensitivity analysis}\label{sec:allocation}
\subsection{Allocation and marginal cost}\label{subsec:allocation}

The objective gains at rate $c_j$ as $d_j$ increases, while the budget charges $a_j C(d_j)$. The ratio $c_j/a_j$ therefore governs the allocation. The following cost and displacement functions reduce the optimization to a scalar equation for $t\ge 0$:
\[
 \beta(t)=\begin{cases}
 t^2/3,& 0\le t\le 1,\\[3pt]
 1-\dfrac{2}{3t},& t>1,
 \end{cases}
 \qquad
 G(t)=\begin{cases}
 t/3,& 0\le t\le 1,\\[3pt]
 \dfrac{1}{2}-\dfrac{1}{6t^2},& t>1.
 \end{cases}
\]
With $\act=\{j:c_j>0\}$, define the saturation threshold $\sat=\sum_{j\in\act}a_j$, the benefit-to-weight ratios $z_j=c_j/a_j$ ($Z=\max_{1\le j\le J}z_j$), and the quadratic scale $S=\sum_{j=1}^J a_j z_j^2=\sum_{j=1}^J c_j^2/a_j$. For a scalar multiplier proxy $h\ge 0$, set $\Phi_B(h)=\sum_{j=1}^J a_j \beta(h z_j)$ and $\Phi_R(h)=\sum_{j=1}^J c_j G(h z_j)$.

\begin{theorem}\label{thm:B1}
Fix $\boldsymbol{p},\boldsymbol{q},\boldsymbol{a}\in\Delta_J$ with $a_j>0$ for all $j=1,\ldots,J$.
\begin{enumerate}
\item If $\act=\emptyset$, then $R(r)=0$ for all $r\in[0,1]$.
\item If $\act\ne\emptyset$, then $R(0)=0$, $R(r)=R_{\max}$ for all $r\in[\sat,1]$, and for each $r\in(0,\sat)$ the equation $\Phi_B(h)=r$ has a unique root $h>0$, the entire optimal vector is unique, with $d_j^*=G(hz_j)$ for every $j$ (zero off $\act$), and $R(r)=\Phi_R(h)$. Moreover, $R$ is continuous on $[0,1]$ and strictly increasing and strictly concave on $[0,\sat]$.
\item If $\act\ne\emptyset$, put $r_0=S/(3Z^2)\le \sat/3$. Then $R(r)=\sqrt{rS/3}$ for $0\le r\le r_0$, and $R(r)<\sqrt{rS/3}$ for all $r\in(r_0,1]$.
\end{enumerate}
\end{theorem}

\begin{remark}
For $\act\ne\emptyset$, the quadratic Cauchy--Schwarz bound is exact for $r\le r_0$. Above $r_0$, bins with the largest collision-to-weight ratios are the first to enter the higher-cost regime. The full cost gives a strictly narrower identification interval than the quadratic relaxation.
\end{remark}

For an interior budget, the active coordinates satisfy
\[
 c_j=\lambda a_j C'(d_j^*),\qquad h=1/(2\lambda).
\]
The multiplier $\lambda>0$ equates the marginal AUC gain per unit budget across the overlapping bins. In the quadratic regime, $d_j^*=hz_j/3$, so the ratios $z_j$ determine both the allocation and which bin first leaves that regime. When the budget permits saturation, each remaining collision can be ordered entirely in the adverse direction.

At saturation every overlapping bin reaches $1/2$. Setting inactive coordinates to zero selects a canonical optimizer when the budget exceeds $\sat$. The \hyperlink{proof:thm:B1}{proof of Theorem~\ref*{thm:B1}} in Appendix~\ref*{app:proofs} derives the optimal allocation.

\subsection{Critical budgets}

The lower endpoint answers whether $\theta>\tau$ holds for every compatible population. Following the breakdown perspective of \citet{Kline2013,Masten2020}, define the critical information budget by its first crossing,
\[
 r^*(\boldsymbol{p},\boldsymbol{q},\boldsymbol{a};\tau)=\inf\bigl\{r\in[0,1]:L(\boldsymbol{p},\boldsymbol{q},\boldsymbol{a};r)\le\tau\bigr\}\in[0,1]\cup\{+\infty\},\qquad \inf\emptyset=+\infty,
\]
abbreviated $r^*(\tau)$ at $(\boldsymbol{p},\boldsymbol{q})$. It gives the smallest budget at which the identified set no longer rules out $\theta\le\tau$, quantifying how much unobserved within-bin signal can be tolerated before $\theta>\tau$ loses identification support.

\begin{corollary}\label{cor:C}
For any threshold $\tau\in\mathbb{R}$, with $L_{\min}=m-R_{\max}$,
\[
 r^*(\tau)=\begin{cases}
 0,& \tau\ge m,\\
 +\infty,& \tau<L_{\min},\\
 \sat,& \tau=L_{\min}<m,\\
 \Phi_B(h)\text{ for the unique }h>0\text{ solving }\Phi_R(h)=m-\tau,& L_{\min}<\tau<m.
 \end{cases}
\]
In the last branch, if $0<m-\tau\le S/(3Z)$, then $r^*(\tau)=3(m-\tau)^2/S$. Every finite $r^*(\tau)$ satisfies $L(r^*(\tau))\le\tau$.
\end{corollary}

An infinite critical budget means that even unrestricted within-bin ordering leaves $\theta>\tau$. At $\tau=L_{\min}<m$, the first crossing is at saturation. The \hyperlink{proof:cor:C}{proof of Corollary~\ref*{cor:C}} in Appendix~\ref*{app:proofs} derives the first-crossing formula.

\section{Finite-Sample Inference}\label{sec:inference}

The bin probabilities are estimated by $\hat{\boldsymbol{p}}$ and $\hat{\boldsymbol{q}}$, giving sample endpoints $\hat L(r)=L(\hat{\boldsymbol{p}},\hat{\boldsymbol{q}},\boldsymbol{a};r)$ and $\hat U(r)=U(\hat{\boldsymbol{p}},\hat{\boldsymbol{q}},\boldsymbol{a};r)$. We first quantify uncertainty at a given budget or threshold, then obtain a common confidence event for the full curves. A separate test addresses the decision at a prespecified pair $(r,\tau)$. The methods answer different questions. Pointwise inference targets a single identified set or threshold crossing. Simultaneous bands cover the set-valued sensitivity curve on a common event and allow the reader to inspect multiple thresholds. The fixed-pair test asks whether a particular claim about the lower endpoint can be certified.

\subsection{Pointwise KL inference}\label{subsec:kl}

Put $k=\min(n_0,n_1)$ and let $D(x\Vert y)$ denote the Bernoulli Kullback--Leibler (KL) divergence with standard boundary conventions. For any fixed measurable kernel taking values in $[0,1]$, let $W$ be its two-sample U-statistic average over all $n_0n_1$ pairs and let $\mu=\E W$. Conditioning a random matching of size $k$ on the two samples and applying Jensen's inequality bounds the moment-generating function of $kW$ by that of a $\mathrm{Binomial}(k,\mu)$ variable. Theorem~1 and Section~5b of \citet{Hoeffding1963} give the corresponding tail bounds
\[
 \Prob(W\ge x)\le e^{-kD(x\Vert\mu)}\quad(x\ge\mu),
 \qquad
 \Prob(W\le x)\le e^{-kD(x\Vert\mu)}\quad(x\le\mu)
\]. For $\delta\in(0,1)$, define the inverted KL bounds
\[
 \ell_\delta(x)=\inf\{y\in[0,x]:kD(x\Vert y)\le\log(1/\delta)\},
 \qquad u_\delta(x)=1-\ell_\delta(1-x),
\]
which are continuous and strictly increasing on $[0,1]$ with $\ell_\delta(0)=0$ and $\ell_\delta(1)=\delta^{1/k}$.

\begin{proposition}\label{prop:kl}
Fix $\alpha\in(0,1)$ and define $I_{\KL}(r)=[\ell_{\alpha/2}(\hat L(r)),u_{\alpha/2}(\hat U(r))]$.
\begin{enumerate}
\item For each prespecified $r\in[0,1]$, $\Prob([L(r),U(r)]\subseteq I_{\KL}(r))\ge 1-\alpha$.
\item Under \hyperlink{assumption:M1}{(M1)}--\hyperlink{assumption:M4}{(M4)}, a single event of probability at least $1-\alpha$ guarantees $\theta\in I_{\KL}(r)$ simultaneously for every $r\ge\rho$.
\item For a prespecified threshold $\tau\in\mathbb{R}$ and level $\delta\in(0,1)$, $\underline r_{\KL,\delta}(\tau)=\inf\{r\in[0,1]:\ell_\delta(\hat L(r))\le\tau\}$ ($\inf\emptyset=+\infty$) satisfies $\Prob(\underline r_{\KL,\delta}(\tau)\le r^*(\tau))\ge 1-\delta$.
\end{enumerate}
\end{proposition}

At a fixed budget the population minimizer and maximizer are also feasible for the sample optimization. Their sample averages therefore bound $\hat L(r)$ from above and $\hat U(r)$ from below, so paired KL can be applied to two fixed kernels. For actual-AUC coverage, the true win vector supplies a single fixed kernel feasible at every $r\ge\rho$. For the critical budget, the fixed witness is the population minimizer at the first crossing, whenever that crossing is finite. Monotonicity of $\ell_\delta$ then reduces computation of the sample limit to Corollary~\ref{cor:C} with inverted threshold $x_\delta(\tau)$. The \hyperlink{proof:prop:kl}{proof of Proposition~\ref*{prop:kl}} in Appendix~\ref*{app:proofs} establishes pointwise coverage.

\subsection{Simultaneous confidence bands}\label{subsec:bands}

A confidence envelope for the bin distributions can support inference at every budget on a single event. Let $F_g$ and $\hat F_g$ be the population and empirical bin cumulative distribution functions (CDFs) for group $g\in\{0,1\}$. At internal cuts $j=1,\ldots,J-1$, define clipped DKWM envelopes $F_g^\pm(j)=\clip(\hat F_g(j)\pm\varepsilon_g,0,1)$ with $\varepsilon_g=\sqrt{\log(4/\alpha)/(2n_g)}$, keeping $F_g^\pm(0)=0$ and $F_g^\pm(J)=1$. Differencing gives mass vectors $\boldsymbol{p}^\pm,\boldsymbol{q}^\pm\in\Delta_J$, and the joint two-sample event $\{F_g^-\le F_g\le F_g^+,\ g=0,1\}$ has probability at least $(1-\alpha/2)^2>1-\alpha$ \citep{Massart1990,LeBoudec2026}.

The CDF envelopes translate into endpoint bounds because the kernel is monotone in the two bin indices. Summation by parts makes this relation explicit. For fixed $\boldsymbol{v}\in[0,1]^J$ and marginals $\boldsymbol{p}',\tilde{\boldsymbol{p}},\boldsymbol{q}',\tilde{\boldsymbol{q}}\in\Delta_J$, setting $\xi_i=(\mathbf{K}_{\boldsymbol{v}}\boldsymbol{q}')_i$ and $\zeta_j=(\mathbf{K}_{\boldsymbol{v}}^{\mathsf T}\tilde{\boldsymbol{p}})_j$ gives
\[
 T(\boldsymbol{p}',\boldsymbol{q}';\boldsymbol{v})-T(\tilde{\boldsymbol{p}},\tilde{\boldsymbol{q}};\boldsymbol{v})
 =\sum_{i=1}^{J-1}\bigl(F_{\boldsymbol{p}'}(i)-F_{\tilde{\boldsymbol{p}}}(i)\bigr)(\xi_i-\xi_{i+1})
 +\sum_{j=1}^{J-1}\bigl(F_{\tilde{\boldsymbol{q}}}(j)-F_{\boldsymbol{q}'}(j)\bigr)(\zeta_{j+1}-\zeta_j),
\]
where $\xi_i-\xi_{i+1}=q'_iv_i+q'_{i+1}(1-v_{i+1})\ge 0$ and $\zeta_{j+1}-\zeta_j=\tilde p_j(1-v_j)+\tilde p_{j+1}v_{j+1}\ge 0$. Because the feasibility set $V_r$ depends only on $(\boldsymbol{a},r)$ and not on $(\boldsymbol{p}',\boldsymbol{q}')$, taking infima and suprema over $V_r$ preserves stochastic-order monotonicity for every $r\in[0,1]$.

\begin{proposition}\label{prop:band}
For the DKW or Clopper--Pearson (CP) CDF envelopes $(F_g^-,F_g^+)$, put $L_b(r)=L(\boldsymbol{p}^-,\boldsymbol{q}^+,\boldsymbol{a};r)$, $U_b(r)=U(\boldsymbol{p}^+,\boldsymbol{q}^-,\boldsymbol{a};r)$, and $\underline r_b(\tau)=\inf\{r\in[0,1]:L_b(r)\le\tau\}$ ($\inf\emptyset=+\infty$). Then
\[
 \Prob\!\left(
 \bigl\{[L(r),U(r)]\subseteq[L_b(r),U_b(r)]\ \forall r\in[0,1]\bigr\}
 \cap\bigl\{\underline r_b(\tau)\le r^*(\tau)\ \forall\tau\in\mathbb R\bigr\}
 \right)\ge 1-\alpha .
\]
For DKW, the joint coverage is at least $(1-\alpha/2)^2$, and the frontier event alone holds with probability at least $(1-\alpha/4)^2$ using the same radii $\varepsilon_g$.
\end{proposition}

An alternative envelope builds Clopper--Pearson (CP) binomial intervals \citep{ClopperPearson1934} for each cumulative count at internal cuts $j=1,\ldots,J-1$ (for $J>1$) with tail allocation $\alpha/[4(J-1)]$, taking prefix maxima of lower bounds and suffix minima of upper bounds to ensure monotonicity. On the CDF confidence event, both population marginals lie between their corresponding ordered corners. The corner endpoints are attainable within the CDF box. All intervening values are also attained because the product of the two CDF boxes and the feasible win-vector set is connected and its image under the AUC map is continuous. Opposite-corner projection $(\boldsymbol{p}^-,\boldsymbol{q}^+)$ and $(\boldsymbol{p}^+,\boldsymbol{q}^-)$ achieves the exact union of sharp intervals over the entire CDF box. The \hyperlink{proof:prop:band}{proof of Proposition~\ref*{prop:band}} in Appendix~\ref*{app:proofs} justifies the projection and simultaneous coverage.

\subsection{Certified fixed-pair testing}\label{subsec:cert}

At a prespecified $(r,\tau)$, testing $H_0:L_\vartheta\le\tau$ can use the variance of the kernel that attains the lower endpoint, where $\vartheta=(\boldsymbol{p},\boldsymbol{q})$ and $L_\vartheta=L(\boldsymbol{p},\boldsymbol{q},\boldsymbol{a};r)$. Let $\mathbf{K}_\vartheta$ be the minimizing kernel from Theorem~\ref{thm:B1}, with diagonal $1/2-d_j^*$ on $\act$ and $1/2$ off $\act$. For fixed $\vartheta$ it is a fixed kernel, and sample minimization gives $\hat L(r)\le\hat{\boldsymbol{p}}^{\mathsf T}\mathbf{K}_\vartheta\hat{\boldsymbol{q}}$. Thus a large empirical gap above $\tau$ under the null requires a large fluctuation of this kernel average. With $R_i,S_j$ its row and column means, let $f_i=R_i-L_\vartheta$, $g_j=S_j-L_\vartheta$, and $D_{ij}=\mathbf{K}_\vartheta(i,j)-R_i-S_j+L_\vartheta$. The Hoeffding decomposition of the fluctuation is $P_\vartheta+W_\vartheta$, where
\[
 P_\vartheta=\frac{1}{n_0}\sum_{s=1}^{n_0}f_{B(X_s)}+\frac{1}{n_1}\sum_{t=1}^{n_1}g_{B(Y_t)},\qquad
 W_\vartheta=\frac{1}{n_0n_1}\sum_{s=1}^{n_0}\sum_{t=1}^{n_1}D_{B(X_s),B(Y_t)}.
\]
The $n_0+n_1$ summands $Z_\ell$ in $P_\vartheta$ are independent with mean zero, giving variance $V=\sum_\ell\E Z_\ell^2$, third moment $\kappa=\sum_\ell\E Z_\ell^3$, and sum of fourth increment moments $m_4=\sum_\ell\E Z_\ell^4$. The fourth moment of $P_\vartheta$ also contains cross-variance terms. The remainder has variance $v_D=\E D_{B(X_1),B(Y_1)}^2/(n_0n_1)$, so Cantelli's inequality bounds $\Prob_\vartheta(W_\vartheta>e_\vartheta)\le\delta_D$ at $e_\vartheta=\sqrt{v_D(1-\delta_D)/\delta_D}$, which also holds when $v_D=0$.

Following confidence-set localization \citep{BergerBoos1994}, we intersect individual coordinate CP intervals (at two-sided error $\gamma/(2J)$ per marginal bin) with $\Delta_J\times\Delta_J$ to form a root nuisance box $\mathcal{C}$ of noncoverage at most $\gamma$. These intervals localize the bin probabilities that enter the projection moments. Over any sub-box $B\subseteq\mathcal{C}$, Appendix~\ref{app:probability-boxes} encloses the optimal allocation $d_j^*$, lower endpoint $L_\vartheta$, projection moments $(V,\kappa,m_4)$, maximum summand magnitude, and $v_D$, producing an upper bound $e_B$ on $e_\vartheta$ and a certified lower bound on the projection threshold $x_B\le\hat L(r)-\tau-e_B$.

To see how the envelope uses the optimizer, let $\boldsymbol c^-$ and $\boldsymbol c^+$ bound the collision masses in a box. The inverse budget scale is nonincreasing as the collision masses increase, while an individual displacement is nondecreasing with both the scale and its own collision mass. Lower displacement bounds therefore combine the scale at $\boldsymbol c^+$ with the coordinate at $c_j^-$, with the reverse choice for the upper bounds. The upper bounds also use the single-bin budget cap. These displacement intervals give entrywise kernel bounds. Bounds on row and column means then follow from linear optimization over the respective simplex boxes. Centering and taking outward bounds on powers gives the common projection moment enclosure, including the signed third moment. Appendices~\ref{app:probability-boxes} and~\ref{app:numerical-cert} give the boundary conventions and directed-rounding rules.

Write $V_+$ and $b_+$ for the box upper bounds on $V$ and $\max_\ell|Z_\ell|$. For $x>0$ and $V_+,b_+>0$, the mathematical tail bound on $P_\vartheta$ combines Bennett's inequality \citep{Bennett1962} with a Vaaler--Prawitz smoothed Fourier inversion bound \citep{Prawitz1972,Vaaler1985,HallWang2004}:
\begin{align*}
 \Prob_\vartheta(P_\vartheta\ge x)&\le \min\{1,B_B(x),Q_B(x)\},\\
 B_B(x)&=\exp\!\left[-\frac{V_+}{b_+^2}\,h\!\left(\frac{b_+x}{V_+}\right)\right],
 \qquad h(z)=(1+z)\log(1+z)-z .
\end{align*}
Here $Q_B(x)=A_1+A_2+A_{3,\mathrm{lo}}+A_{3,\mathrm{hi}}$ when variance lower-bound and frequency-cutoff admissibility conditions hold, and $Q_B(x)=1$ otherwise. Appendix~\ref{app:tail-bounds} derives these terms: $A_1$ encloses the Gaussian smoothing tail, $A_2$ encloses the signed third-moment correction, and $A_{3,\mathrm{lo}},A_{3,\mathrm{hi}}$ bound characteristic-function discrepancy across low and high frequencies while preserving the boundary atom at $x$. When $V_+=0$ or $b_+=0$, the bound equals $1$ for $x\le 0$ and $0$ for $x>0$.

The variance and third moment enter the smoothed normal approximation, and the remaining terms bound its discrepancy with the characteristic function of the projection. The tail uses an inclusive comparison to retain the atom at the threshold. Bennett supplies a bound when the Fourier admissibility or verification checks fail, and the search accepts a tail certificate only after comparing a verified upper bound with the allocated error probability.

\begingroup
\IncMargin{1.5em}
\begin{algorithm2e}[H]
\caption{Cert-Prawitz-BnB at a prespecified $(r,\tau)$}\label{alg:cert}
\If{the initial empirical gap is nonpositive}{return non-rejection\;}
Evaluate the root CP box. Remove any evaluated box certified empty, outside the null, or to meet the projection tail cutoff at $x_B$\;
\If{an unresolved box triggers the center stop}{return non-rejection\;}
\While{the queue is nonempty and fewer than 256 splits have been used}{
 Select the unresolved box with largest recorded tail score\;
 \If{its widest coordinate is narrower than $10^{-5}$}{return non-rejection\;}
 Bisect that coordinate into two closed children and evaluate both\;
 \If{an unresolved child triggers the center stop}{return non-rejection\;}
 Retain the children that have not been certified\;
}
Reject if the queue is empty. Otherwise return non-rejection\;
\end{algorithm2e}
\endgroup

Algorithm~\ref{alg:cert} for Cert-Prawitz-BnB searches $\mathcal{C}$ by best-first bisection up to 256 splits, pruning any sub-box that is empty, has a certified lower bound on $L_\vartheta$ exceeding $\tau$, or has a certified projection tail bound at $x_B$ at most $\delta_P$, and returning non-rejection immediately if an unresolved box's center fails to clear the cutoff. The numerical implementation checks CP endpoints with directed binomial tails, using outward endpoints when verification fails, encloses $\hat L(r)$ from below and $e_\vartheta$ from above, rounds differences downward for $x_B$, and evaluates Bennett and admissible Vaaler--Prawitz bounds using IEEE binary64 directed rounding and MPFR/GMP arithmetic (Appendix~\ref{app:numerical-cert}, especially Table~\ref{tab:certification-limits}).

The next result combines the box certificates with the confidence-region and remainder events to control the overall rejection probability.

\begin{theorem}\label{thm:level}
For a prespecified $(r,\tau)$, under \hyperlink{assumption:M1}{(M1)}--\hyperlink{assumption:M4}{(M4)} and the outward-enclosure conditions in Appendices~\ref{app:probability-boxes}--\ref{app:numerical-cert}, Cert-Prawitz-BnB has finite-sample level at most $0.025$ over $H_0:L_\vartheta\le\tau$, with error budget $(\gamma,\delta_D,\delta_P)=(0.003,0.004,0.018)$.
\end{theorem}

On the event $\{\vartheta\in\mathcal{C}\}\cap\{W_\vartheta\le e_\vartheta\}$, rejection under the null requires a tail certificate for a removed box $B$ containing $\vartheta$. The witness inequality gives $x_B\le P_\vartheta$, so $s_\vartheta(P_\vartheta)\le s_\vartheta(x_B)\le\delta_P$, where $s_\vartheta(x)=\Prob_\vartheta(P_\vartheta\ge x)$. The inclusive tail at $P_\vartheta$ is super-uniform. The rejection event is therefore contained in the union of the two excluded events and this fixed-parameter tail event, giving level $\gamma+\delta_D+\delta_P=0.025$. The \hyperlink{proof:thm:level}{proof of Theorem~\ref*{thm:level}} in Appendix~\ref*{app:proofs} establishes finite-sample level control.

\FloatBarrier
\section{Simulation Studies}\label{sec:simulation}

\subsection{Simulation settings}\label{subsec:sim-settings}

The first study asks how the budget affects identification width, coverage, and critical-budget inference. The second uses fresh samples to assess the level and power of the fixed-pair test. Both studies use four ordered bins and 5,000 independent multinomial repetitions per design, with competing methods applied to the same count tables.

The coverage and width study comprises 35,000 repetitions across seven designs covering balanced, imbalanced, sparse, low-overlap, single-bin, continuous Beta ($\mathrm{Beta}(2,5)$ controls versus $\mathrm{Beta}(4,3)$ cases on quarter-point cuts), and oscillatory (uniform controls versus density $1+0.8\sin(8\pi x)$, matching bin masses in both groups) score distributions. In the five discrete-conditional designs, within-bin densities follow the extremal forms $1\mp\clip(t_j(2u-1),-1,1)$. Across grids of budgets $r\in[0,1]$ and thresholds $\tau$, this study evaluates sharp versus relaxed identification widths, fixed-budget identified-set coverage, actual-AUC coverage, simultaneous band coverage, and critical-budget lower confidence limits.

The confirmatory fixed-pair study draws 70,000 fresh repetitions across 14 fixed-$(r,\tau)$ designs comprising six alternatives and eight nulls to benchmark Cert-Prawitz-BnB against paired KL, Bennett, and asymptotic endpoint Wald tests at nominal one-sided level $0.025$. All designs fix $r=0.04$ except a sparse design evaluated at saturation. Null stress tests place thresholds at stored upper enclosures of population lower endpoints, test structural zeros and oscillatory equal-marginals, and include a rare-boundary configuration where one group concentrates in a single bin.

\subsection{Identification and interval estimation}\label{subsec:sim-width-cov}

Figure~\ref{fig:sim-width} contrasts population identification widths and mean 95\% paired KL interval widths across $r\in[0,1]$ in the Balanced and Sparse designs. In the Balanced design at $r=0.04$, the sharp identified interval has width $0.108$ compared with $0.330$ without a budget, reducing the mean 95\% paired KL interval width from $0.401$ to $0.185$. Throughout the quadratic allocation regime $r\le r_0$ of Theorem~\ref{thm:B1}, the sharp bound coincides with the Cauchy relaxation replacing $C(d)$ by $3d^2$, while at $r=0.40$ it improves slightly on Cauchy, lowering the population width from $0.306$ to $0.296$ and the mean KL width from $0.378$ to $0.368$.

\ifreviewmode
\begin{figure}[tbp]
\centering
\includegraphics[width=0.95\linewidth]{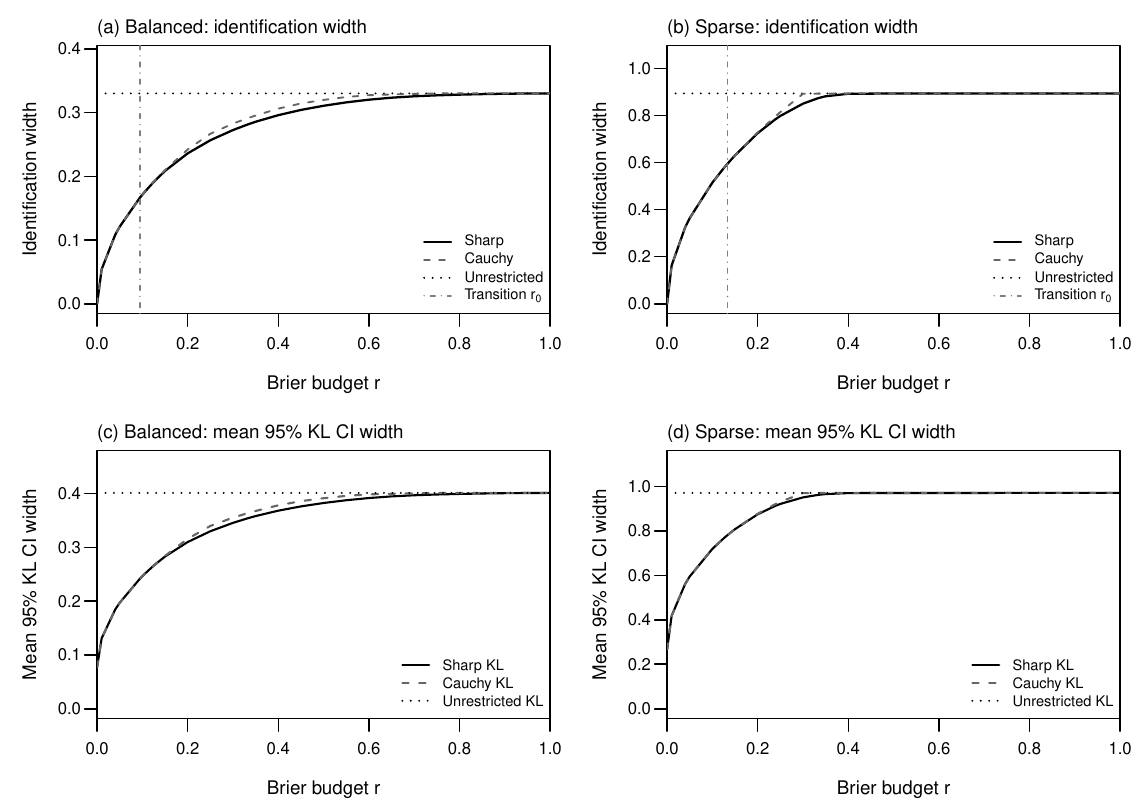}
\caption{Population identification widths (a,~b) and mean 95\% paired KL interval widths (c,~d) in the Balanced and Sparse designs (5,000 replications). Vertical dot-dashed lines mark $r_0=S/(3Z^2)$.}
\label{fig:sim-width}
\end{figure}

\begin{table}[tbp]
\centering
\begin{threeparttable}
\caption{Identified-set coverage (\%) and mean 95\% confidence interval width in the original simulation.}\label{tab:sim-cov-width}
\footnotesize
\setlength{\tabcolsep}{4pt}
\begin{tabular}{lrrrrrrr}
\toprule
Design & $r$ & Sharp KL & Cauchy KL & Unrest. KL & DKW & CP & Endp. Wald \\
\midrule
Balanced & 0.04 & 99.9 / 0.185 & 99.9 / 0.185 & 100.0 / 0.401 & 100.0 / 0.231 & 100.0 / 0.204 & 95.6 / 0.151 \\
 & 0.40 & 99.9 / 0.368 & 100.0 / 0.378 & 100.0 / 0.401 & 100.0 / 0.418 & 100.0 / 0.390 & 95.7 / 0.340 \\
\addlinespace[3pt]
Beta & 0.04 & 100.0 / 0.096 & 100.0 / 0.096 & 100.0 / 0.228 & 100.0 / 0.136 & 100.0 / 0.104 & 94.5 / 0.081 \\
 & 0.40 & 99.9 / 0.210 & 100.0 / 0.216 & 100.0 / 0.228 & 100.0 / 0.249 & 100.0 / 0.219 & 94.8 / 0.197 \\
\addlinespace[3pt]
Imbalanced & 0.04 & 100.0 / 0.343 & 100.0 / 0.343 & 100.0 / 0.543 & 100.0 / 0.338 & 100.0 / 0.292 & 94.3 / 0.204 \\
 & 0.40 & 100.0 / 0.514 & 100.0 / 0.522 & 100.0 / 0.543 & 100.0 / 0.521 & 100.0 / 0.474 & 94.9 / 0.392 \\
\addlinespace[3pt]
Sparse & 0.04 & 100.0 / 0.562 & 100.0 / 0.562 & 100.0 / 0.970 & 100.0 / 0.443 & 100.0 / 0.392 & 95.7 / 0.364 \\
 & 0.40 & 99.6 / 0.970 & 99.6 / 0.970 & 99.7 / 0.970 & 100.0 / 0.972 & 99.5 / 0.941 & 97.7 / 0.935 \\
\bottomrule
\end{tabular}
\begin{tablenotes}
\item \textit{Note:} Entries report coverage / width across 5,000 replications per design. Sharp KL, Cauchy KL, unrestricted KL (Unrest.\ KL), DKW, and CP carry finite-sample guarantees; endpoint Wald (Endp.\ Wald) is asymptotic.
\end{tablenotes}
\end{threeparttable}
\end{table}

\fi
\placement{Figure}{fig:sim-width}
\placement{Table}{tab:sim-cov-width}

Table~\ref{tab:sim-cov-width} summarizes fixed-budget identified-set coverage and mean 95\% interval width across four representative designs at $r=0.04$ and $r=0.40$. Sharp KL covers the identified set in at least $99.58\%$ of repetitions in every assessed budget cell, DKW covers both endpoint curves simultaneously in every repetition, and the lowest simultaneous CP curve coverage is $99.50\%$ in the Sparse design. Simultaneous CP and DKW bands can be narrower than pointwise KL intervals when group sizes are highly unequal or sparse, such as $0.392$ and $0.443$ versus $0.562$ in Sparse at $r=0.04$, whereas asymptotic endpoint Wald dips slightly below $95\%$ in the Beta ($94.5\%$) and Imbalanced ($94.3\%$) designs. The prespecified Holm audit flagged none of its 1,196 coverage checks. Actual-AUC coverage answers a different question when the budget excludes the true information $\rho$: for $r=0$, KL covered the actual AUC in $0\%$ of repetitions in the Beta and Single-bin designs and $5.28\%$ in the Oscillatory design. At the Balanced working threshold $\tau=0.50$, however, the single-threshold KL lower limit reached $0.04$ in only $1,118/5,000$ repetitions ($22.36\%$, with one-sided 95\% Monte Carlo lower limit $21.39\%$), compared with $0.16\%$ for DKW and $3.80\%$ for CP, falling short of a 50\% practical certification target and motivating Section~\ref{subsec:cert}.

\subsection{Type I error, power, and computational performance}\label{subsec:sim-power-null}

Figure~\ref{fig:sim-null} displays empirical rejection rates and 95\% Bonferroni-simultaneous one-sided Monte Carlo upper limits across all 32 method--null cells in the confirmatory study. The observed rejection rates of paired KL, Bennett, and Cert-Prawitz-BnB were below $2.5\%$ in all eight null designs. The highest rate for Cert-Prawitz-BnB was $0.64\%$ under the Beta boundary null, with a simultaneous Monte Carlo upper limit of $1.049\%$. Wald exceeded $2.5\%$ in four null designs. Under the Rare boundary null, the Wald test rejected in 564 of 5,000 repetitions (11.28\%). The estimated standard error was zero in each of those repetitions.

\ifreviewmode
\begin{figure}[tbp]
\centering
\includegraphics[width=0.95\linewidth]{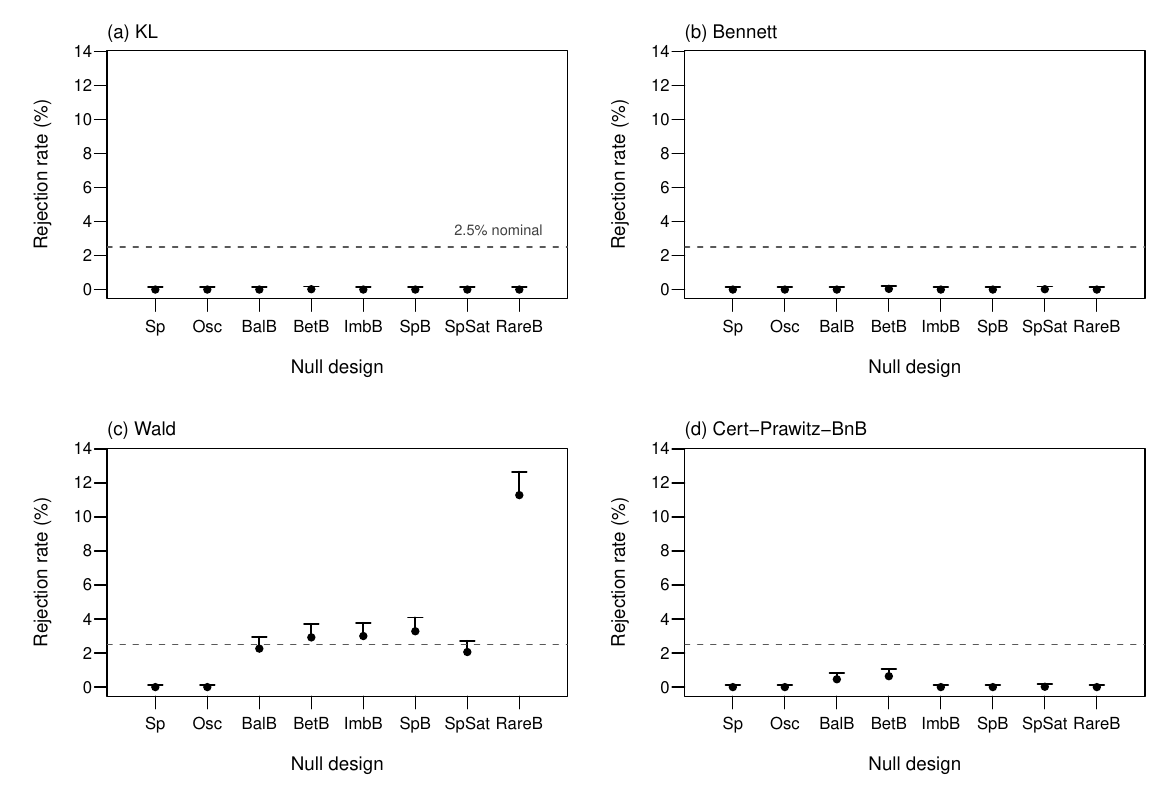}
\caption{Empirical rejection percentages (dots) and 95\% Bonferroni-simultaneous Monte Carlo upper limits (bars) across eight confirmatory nulls for (a)~KL, (b)~Bennett, (c)~Wald, and (d)~Cert-Prawitz-BnB. Each design uses 5,000 repetitions. The dashed line marks 2.5\%. The horizontal axis lists Sparse (Sp), Oscillatory (Osc), Balanced boundary (BalB), Beta boundary (BetB), Imbalanced boundary (ImbB), Sparse boundary (SpB), Sparse at saturation (SpSat), and Rare boundary (RareB).}
\label{fig:sim-null}
\end{figure}

\begin{table}[tbp]
\centering
\begin{threeparttable}
\caption{Confirmatory rejection percentages across the six alternative designs.}\label{tab:sim-main}
\scriptsize
\setlength{\tabcolsep}{2.5pt}
\begin{tabular}{lcrrrrrr}
\toprule
Design & $(n_0, n_1)$ & $r$ & $\tau$ & KL & Bennett & Wald & Cert-Prawitz-BnB \\
\midrule
Balanced & (1200, 1200) & 0.04 & 0.50 & 23.86 [22.7, 25.1] & 28.24 [27.0, 29.5] & 78.12 [76.9, 79.3] & 58.46 [57.1, 59.8] \\
Beta & (4000, 4000) & 0.04 & 0.80 & 13.38 [12.4, 14.4] & 25.70 [24.5, 26.9] & 76.60 [75.4, 77.8] & 57.40 [56.0, 58.8] \\
Imbalanced & (120, 2400) & 0.04 & 0.50 & 0.00 [0.0, 0.1] & 0.28 [0.2, 0.5] & 31.40 [30.1, 32.7] & 1.54 [1.2, 1.9] \\
Local alt. & (1200, 1200) & 0.04 & 0.5196 & 0.58 [0.4, 0.8] & 0.94 [0.7, 1.2] & 16.44 [15.4, 17.5] & 5.76 [5.1, 6.4] \\
Low overlap & (500, 500) & 0.04 & 0.80 & 100.00 [99.9, 100.0] & 100.00 [99.9, 100.0] & 100.00 [99.9, 100.0] & 100.00 [99.9, 100.0] \\
Single bin & (60, 60) & 0.04 & 0.25 & 0.00 [0.0, 0.1] & 0.00 [0.0, 0.1] & 100.00 [99.9, 100.0] & 0.00 [0.0, 0.1] \\
\bottomrule
\end{tabular}
\begin{tablenotes}
\item \textit{Note:} Based on 5,000 independent replications per design at nominal one-sided level 0.025, with two-sided 95\% exact Monte Carlo intervals in brackets. Wald is asymptotic.
\end{tablenotes}
\end{threeparttable}
\end{table}

\fi
\placement{Figure}{fig:sim-null}
\placement{Table}{tab:sim-main}

Table~\ref{tab:sim-main} reports rejection rates and 95\% Monte Carlo confidence intervals across the six confirmatory alternatives. In the moderate-sample Balanced and Beta designs, Cert-Prawitz-BnB raises power to $58.46\%$ and $57.40\%$, respectively. These are gains of $34.60$ and $44.02$ percentage points over paired KL and $30.22$ and $31.70$ points over Bennett. The paired Monte Carlo standard errors for the gains over KL are $0.67$ and $0.70$ points, respectively. The one-sided 95\% Monte Carlo lower limit in the Balanced design is $57.30\%$, above the $50\%$ benchmark. Power falls to $1.54\%$ under severe imbalance and to $5.76\%$ at the local alternative. All methods reach $100\%$ power in the low-overlap design. In the degenerate Single-bin alternative, $L(r)=0.269060>\tau=0.25$, and the finite-sample tests have $0\%$ power while Wald has $100\%$.

Across all 70,000 confirmatory repetitions, Cert-Prawitz-BnB recorded zero warnings, zero numerical failures, and zero CP endpoint verification fallbacks. Runtime varied across repetitions, with a median of $6.28\text{ ms}$, mean of $74.52\text{ ms}$, and 95th percentile of $206.96\text{ ms}$. For Bennett the median, mean, and 95th percentile were $6.62$, $46.36$, and $193.36$ ms. The corresponding times for paired KL were $0.66$, $0.97$, and $2.71$ ms.

\FloatBarrier
\section{Real Data Applications}\label{sec:application}

\subsection{Age discrimination in the PTB-XL cohort}\label{subsec:ptbxl}

Using PTB-XL version~1.0.3 \citep{Wagner2020,PTBXLv103}, we ask whether age discriminates between patients whose ECG reports carry a NORM label and those with a myocardial infarction (MI) label. We retain the record with the smallest ECG identifier per patient, exclude missing ages, and use diagnostic superclass labels with positive likelihood. The resulting groups contain $n_0=8,318$ patients with only NORM and $n_1=4,453$ with MI and no NORM ($N=12,771$). We divide age at $40,50,60,70$, and $80$ years into six equal-weight bins ($a_j=1/6$), assigning privacy-recoded ages $\ge90$ to the upper bin.

The target is the probability that a patient from the MI report group is older than a patient from the NORM report group. The model takes the examination ages to be atomless, puts each latent age in the bin assigned from the recorded integer age, and assumes independent iid sampling within the two groups. We selected PTB-XL using cell sparsity, variance bounds, and the sample lower endpoint at $r=0.04$. The working pair $(r,\tau)=(0.04,0.70)$ was fixed before the method comparison. Because screening used features of the same data, the comparison illustrates inference under the working design and does not account for selection. The threshold is a working benchmark for the ordering comparison. The budget is a sensitivity setting. The release does not specify rounding at the cutpoints, so the link from recorded age bins to latent age is a modeling assumption.

\ifreviewmode
\begin{figure}[tbp]
\centering
\includegraphics[width=0.96\linewidth]{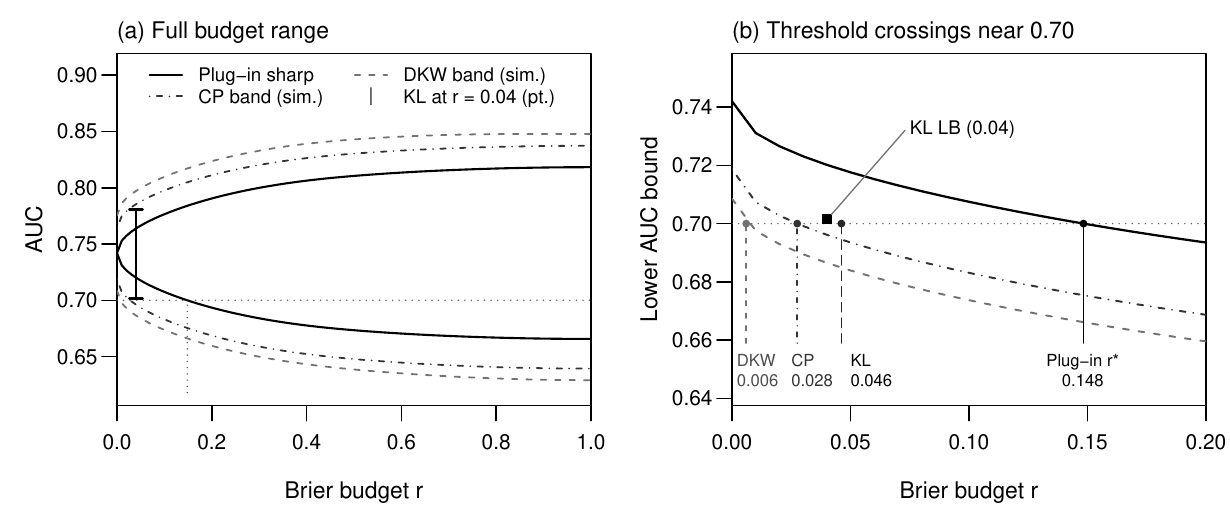}
\caption{Sensitivity of latent-age AUC in PTB-XL ($N=12{,}771$): (a)~full budget range comparing sample sharp bounds with 95\% simultaneous CP/DKW bands and the pointwise KL interval at $r=0.04$; (b)~low-budget crossings with $\tau=0.70$ (dotted line).}
\label{fig:ptb-sensitivity}
\end{figure}

\begin{table}[tbp]
\centering
\begin{threeparttable}
\caption{Exploratory PTB-XL grid scan across six budgets and 121 thresholds.}\label{tab:ptb-scan}
\footnotesize
\setlength{\tabcolsep}{4pt}
\begin{tabular}{lrrrrrrr}
\toprule
$r$ & $\hat L(r)$ & KL & Bennett & Cert-Prawitz-BnB & Wald & vs KL & vs Ben. \\
\midrule
0.00 & 0.7420 & 0.7225 & 0.7275 & 0.7300 & 0.7325 & +3/-0 & +1/-0 \\
0.04 & 0.7202 & 0.7000 & 0.7050 & 0.7075 & 0.7100 & +3/-0 & +1/-0 \\
0.10 & 0.7075 & 0.6875 & 0.6900 & 0.6950 & 0.6975 & +3/-0 & +2/-0 \\
0.25 & 0.6883 & 0.6675 & 0.6725 & 0.6750 & 0.6775 & +3/-0 & +1/-0 \\
0.50 & 0.6734 & 0.6525 & 0.6575 & 0.6600 & 0.6625 & +3/-0 & +1/-0 \\
1.00 & 0.6657 & 0.6450 & 0.6475 & 0.6525 & 0.6550 & +3/-0 & +2/-0 \\
\addlinespace[3pt]
All 726 pairs & --- & 436 & 446 & 454 & 460 & +18/-0 & +8/-0 \\
\bottomrule
\end{tabular}
\begin{tablenotes}
\item \textit{Note:} $\hat L(r)$ is the sample sharp lower bound; method columns report the largest rejected grid threshold per budget (0.50 to 0.80 in steps of 0.0025; total certified pairs in the bottom row). Final columns give Cert-Prawitz-BnB gains/losses relative to KL and Bennett. Grid maxima summarize pointwise exploratory decisions rather than simultaneous lower confidence limits; Wald is asymptotic.
\end{tablenotes}
\end{threeparttable}
\end{table}

\fi
\placement{Figure}{fig:ptb-sensitivity}
\placement{Table}{tab:ptb-scan}

Figure~\ref{fig:ptb-sensitivity} shows how the sample bounds change with the budget. With no restriction on within-bin discrimination ($r=1$), the sample sharp interval $[0.6657,0.8183]$ straddles the working threshold of $0.70$. The grouped mid-rank AUC is $0.7420$, and at $r=0.04$ the interval narrows to $[0.7202,0.7638]$. The 95\% pointwise paired KL interval $[0.7017,0.7808]$ remains above $0.70$ at this budget. The 95\% simultaneous bands include it: $[0.6865,0.7971]$ for DKW and $[0.6962,0.7854]$ for CP. The sample lower curve first reaches $0.70$ at $\hat r^*(0.70)=0.1483$. The one-sided $97.5\%$ paired KL lower confidence limit is $0.0462$; the 95\% simultaneous DKW and CP lower limits are $0.0060$ and $0.0275$, respectively. Figure~\ref{fig:ptb-sensitivity}(b) shows these limits.

At $(r,\tau)=(0.04,0.70)$, all four tests reject $H_0:L(0.04)\le0.70$ at nominal level $0.025$. The certified test uses Bennett at this pair and requires no subdivision of $\mathcal C$. The exploratory comparison in Table~\ref{tab:ptb-scan} shows where Fourier calibration affects decisions. It spans six budgets and 121 thresholds from $0.50$ to $0.80$ in steps of $0.0025$. Of the 726 pairs, the certified test rejects at 454, compared with 436 for KL, 446 for Bennett, and 460 for asymptotic Wald. It adds 18 pairs over KL and 8 over Bennett, retaining every pair certified by either comparator. A Prawitz certificate accounts for each gain over Bennett. The grid maxima in Table~\ref{tab:ptb-scan} summarize these pointwise decisions. They are descriptive summaries of the scan. For statements about multiple thresholds on a common confidence event, the simultaneous bands and their inversions provide the corresponding inferential tool.

\subsection{Grouped age in the esophageal cancer study}\label{subsec:esoph}

The Ille-et-Vilaine case--control study compares hospital cases with controls drawn from electoral rolls \citep{BreslowDay1980,REsoph}. We aggregate over alcohol and tobacco strata to compare the ages of 200 cases and 775 controls in six ordered age groups, from 25--34 to 75 and above. The model assumes independent sampling within and across groups and latent continuous ages consistent with the reported bins. The question is whether the groups can be ordered by age despite the absence of individual ages.

\ifreviewmode
\begin{figure}[tbp]
\centering
\includegraphics[width=0.96\linewidth]{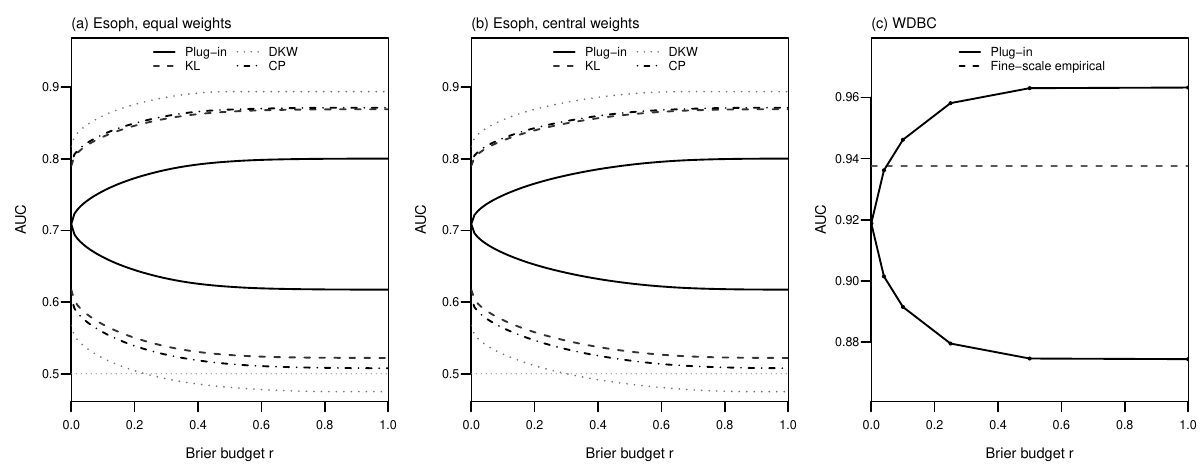}
\caption{Sensitivity bounds for esoph under equal~(a) and central~(b) weights, and for WDBC~(c) against the fine-scale empirical AUC (dashed line).}
\label{fig:supp-app}
\end{figure}

\fi
\placement{Figure}{fig:supp-app}

The grouped mid-rank AUC is $0.7089$, with an unrestricted plug-in interval of $[0.6174,0.8004]$. At $r=0.04$, equal weights give a plug-in interval of $[0.6797,0.7380]$, while weights $(1,1,2,2,2,1)/9$ give $[0.6835,0.7343]$. The central weighting charges more for the same amount of Brier information in the three central age groups. The corresponding 95\% paired KL intervals are $[0.5865,0.8160]$ and $[0.5904,0.8127]$. Figure~\ref{fig:supp-app}(a)--(b) shows the two weighting analyses separately. At $\tau=0.50$, neither plug-in lower curve crosses the benchmark. The KL and CP lower confidence limits are also infinite for both weightings, whereas the DKW limits are $0.2326$ and $0.3002$, respectively. At $(0.04,0.50)$ all four methods reject for both weightings, and the certified test uses Bennett without splitting.

\subsection{Coarsened nuclear radius in the Wisconsin breast cancer data}\label{subsec:wdbc}

The Wisconsin Diagnostic Breast Cancer data contain scores for 357 benign and 212 malignant breast masses evaluated by fine-needle aspiration \citep{WDBCUCI}. We use mean nuclear radius to rank the masses, with larger values ranking toward malignancy. We group the released scores into seven equal-weight bins with cutpoints $10,12,14,16,18,20$. These cutpoints were chosen for display. The count analysis treats the groups as independent samples from distributions of latent continuous scores compatible with the bins.

The grouped mid-rank AUC is $0.9188$. The plug-in interval is $[0.9015,0.9362]$ at $r=0.04$ and $[0.8745,0.9632]$ without a budget. The empirical AUC from the released fine-scale values, with half credit for ties, is $0.9375$. As shown in Figure~\ref{fig:supp-app}(c), it lies slightly above the plug-in upper endpoint at $r=0.04$ and inside the unrestricted interval. This is a descriptive comparison of sample quantities. It illustrates how a working budget can narrow the range enough to exclude a fine-scale empirical reference.

Across 1,158 exploratory pairs spanning both esoph weightings and WDBC, the certified test agreed with Bennett at every pair. In esoph it gained one decision and lost seven against paired KL per weighting. WDBC accounted for 56 further losses and no gains. Together these analyses illustrate that a more elaborate tail calculation can yield the same decisions as Bennett even where the box search is costly.

\FloatBarrier
\section{Discussion}\label{sec:discussion}

The PTB-XL example asks how much hidden age information would make an AUC of $0.70$ compatible with the grouped data. The sample critical budget was $0.1483$. At that budget, the lower endpoint computed from the observed bin proportions first reached the threshold. Near zero budget, bins with larger case--control collision probabilities per unit of budget weight receive larger optimal within-bin AUC departures. Paired KL bounds give a confidence interval for the population identified set at a fixed budget. Projected DKW or Clopper--Pearson bands cover the curves across budgets and give simultaneous lower confidence limits for the population critical budgets.

Cert-Prawitz-BnB tests the lower endpoint against a prespecified threshold at a prespecified budget. Under the stated sampling and numerical conditions, its finite-sample level is at most $0.025$. In the Balanced and Beta simulations, the certified test had higher power than Bennett. In the exploratory PTB-XL scan, Fourier calibration changed some decisions. The certified test agreed with Bennett throughout the esoph and WDBC scans. A tighter projection bound can matter when the decision hinges on the projection tail. A large remainder allowance or substantial uncertainty about bin probabilities can leave little scope for that improvement. The certified search also took longer on average than paired KL. That cost matters when deciding whether to certify a particular threshold comparison.

Future work could use validation data with both original scores and released categories to check the assumed bin mapping. The same data could help calibrate plausible Brier budgets across applications. A second direction would extend the inference to samples with dependence within centres. Such an extension would seek finite-sample guarantees under an explicit model for clustered sampling.

\section*{Acknowledgements}
This research was supported by the Natural Science Foundation of Xinjiang Uygur Autonomous Region (No.~2025D14015) and the National Natural Science Foundation of China (No.~72571102).

\section*{Competing interests}
The authors declare no competing financial interests or personal relationships that could influence the work reported in this paper.

\section*{Data and code availability}
PTB-XL version 1.0.3 is available from PhysioNet \citep{PTBXLv103}. The esophageal cancer data are distributed as \texttt{esoph} in the R \texttt{datasets} package \citep{REsoph}. The Wisconsin Diagnostic Breast Cancer data are available from the UCI Machine Learning Repository \citep{WDBCUCI}. The review package contains the method source, figure and table data, and display scripts. \emph{Permanent archive link pending.}

\clearpage
\begingroup
\small
\setlength{\bibsep}{1pt}
\bibliographystyle{plainnat}
\bibliography{references}
\endgroup
\clearpage
\appendix
\titleformat{\section}{\normalfont\large\bfseries}{\appendixname~\thesection}{0.65em}{}
\counterwithin{equation}{section}
\counterwithin{table}{section}
\counterwithin{figure}{section}
\counterwithin{lemma}{section}
\counterwithin{theorem}{section}
\counterwithin{proposition}{section}
\counterwithin{corollary}{section}
\renewcommand{\theequation}{\Alph{section}.\arabic{equation}}
\renewcommand{\theHequation}{appendix.\Alph{section}.\arabic{equation}}
\renewcommand{\thetable}{\Alph{section}.\arabic{table}}
\renewcommand{\thefigure}{\Alph{section}.\arabic{figure}}
\renewcommand{\thelemma}{\Alph{section}.\arabic{lemma}}
\renewcommand{\thetheorem}{\Alph{section}.\arabic{theorem}}
\renewcommand{\theproposition}{\Alph{section}.\arabic{proposition}}
\renewcommand{\thecorollary}{\Alph{section}.\arabic{corollary}}
\section{Proofs of the main results}\label{app:proofs}

Throughout, $R_{\max}=\frac{1}{2}\sum_{j=1}^J c_j$, $\Phi_B(h)=\sum_{j=1}^J a_j\beta(hz_j)$, and $\Phi_R(h)=\sum_{j=1}^J c_j G(hz_j)$.

\Needspace{4\baselineskip}\begin{proof}[\hypertarget{proof:lem:A}{Proof} of Lemma~\ref{lem:A}]

Suppress the bin index $j\in\act$. Draw a latent group indicator $E\in\{0,1\}$ with $\Prob(E=1)=1/2$ and a score $Z\mid E=g\sim P_g$, so $Z\sim M=(P_0+P_1)/2$ and $\eta(Z)=\Prob(E=1\mid Z)$ satisfies $\E_M[\eta(Z)]=1/2$. Conditioning on $Z$ yields
\[
 \E\bigl[(E-h(Z))^2\bigr]=\E_M\bigl[\eta(1-\eta)\bigr]+\E_M\bigl[(h-\eta)^2\bigr],
\]
which is minimized at $h=\eta$, whereas the constant prediction $h\equiv 1/2$ incurs baseline risk $1/4$. The relative risk reduction is therefore
\[
 4\Bigl(\frac{1}{4}-\E_M[\eta(1-\eta)]\Bigr)=4\bigl(\E_M[\eta^2]-(\E_M\eta)^2\bigr)=4\Var_M(\eta)=b,
\]
matching the quadratic score representation in \citet{Reid2011}.

Because $P_0$ and $P_1$ are atomless, $M$ has a continuous cumulative distribution function $F_M$. Let $Q_M(u)=\inf\{x\in\mathbb{R}:F_M(x)\ge u\}$ for $u\in(0,1)$. Continuity of $F_M$ yields $F_M(Q_M(u))=u$, $Q_M(u)\le x\iff u\le F_M(x)$, $F_M(Z)\sim\mathrm{Unif}(0,1)$, and $Q_M(F_M(Z))=Z$ $M$-almost surely (hence $P_0$- and $P_1$-almost surely). Define $s(u)=2\eta(Q_M(u))-1$ and $w(u)=2u-1$ on $(0,1)$. Since $dP_0/dM=2(1-\eta)$ and $dP_1/dM=2\eta$, the transformed rank variables $U_0=F_M(X)$ and $U_1=F_M(Y)$ (for $X\sim P_0$, $Y\sim P_1$ independent) have densities $1-s$ and $1+s$ on $(0,1)$, with $|s(u)|\le 1$, $\int_0^1 s(u)\,du=0$, and $b=\int_0^1 s(u)^2\,du$. Atomlessness eliminates rank ties, so
\[
 v=\Prob(X<Y)=\int_{0<x<y<1}\bigl(1-s(x)\bigr)\bigl(1+s(y)\bigr)\,dx\,dy.
\]
Using symmetry in $(x,y)$ and $\int_0^1 s(u)\,du=0$, we have $\int_{0<x<y<1}s(x)s(y)\,dx\,dy=\frac{1}{2}(\int_0^1 s)^2=0$. Applying Fubini's theorem to the remaining terms gives
\[
 v-\frac{1}{2}=\int_0^1 w(u)s(u)\,du,\qquad b=\int_0^1 s(u)^2\,du.
\]
Replacing $s$ by $-s$ reverses the sign of $v-1/2$. For any $t>0$, the clipped linear profile $s_t(u)=\clip(tw(u),-1,1)$ maximizes $wz-z^2/(2t)$ uniquely over $z\in[-1,1]$ pointwise and satisfies $\int_0^1 s_t(u)\,du=0$ by antisymmetry about $u=1/2$. Writing $\beta(t)=\int_0^1 s_t(u)^2\,du$ and $G(t)=\int_0^1 w(u)s_t(u)\,du$, pointwise optimality gives
\[
 \int_0^1 w(u)s(u)\,du \le G(t)+\frac{1}{2t}\left(\int_0^1 s(u)^2\,du-\beta(t)\right),
\]
with equality if and only if $s=s_t$ almost everywhere. Since $|w|\sim\mathrm{Unif}(0,1)$, direct integration yields $\beta(t)=t^2/3$ and $G(t)=t/3$ for $0\le t\le 1$, while for $t>1$,
\[
 \beta(t)=\int_0^{1/t} t^2 u^2\,du + \int_{1/t}^1 1\,du = 1-\frac{2}{3t},\qquad
 G(t)=\int_0^{1/t} t u^2\,du + \int_{1/t}^1 u\,du = \frac{1}{2}-\frac{1}{6t^2}.
\]
Solving $G(t)=d\in(0,1/2)$ gives $t=3d$ for $0<d\le 1/3$ and $t=(3-6d)^{-1/2}$ for $1/3<d<1/2$, forcing $b\ge \beta(t)=C(d)$ with equality at $\pm s_t$. At $d=0$, $b\ge 0$ holds with equality at $s\equiv 0$; at $d=1/2$, $|\int_0^1 ws|\le \int_0^1 |w|=1/2$ forces $s=\pm\operatorname{sign}(w)$ almost everywhere and $b=1$. Pushing $1\mp s$ forward onto any finite open subinterval $(l,h)\subset I_j$ via $u\mapsto l+(h-l)u$ produces atomless conditional laws achieving $(\pm d, C(d))$. Finally, continuity at $d=1/3$ together with $C'(d)=6d$ on $(0,1/3]$ and $C'(d)=2/\sqrt{3-6d}$ on $(1/3,1/2)$ (satisfying $C'(1/3-)=C'(1/3+)=2$) establishes strict monotonicity and strict convexity on $[0,1/2]$. Both branches of Lemma~\ref{lem:A} also follow from Theorem~2.2 and Section~3.2 of \citet{Cornilly2018}; the quadratic branch follows from the Gini mean difference bound of \citet{Pham2011}. \end{proof}

\begin{proof}[\hypertarget{proof:thm:B}{Proof} of Theorem~\ref{thm:B}]
Conditioning on the bins gives $\theta=T(\boldsymbol p,\boldsymbol q;\boldsymbol v)$. Lemma~\ref{lem:A} on $\act$ and convention~\hyperlink{assumption:M3}{(M3)} on its complement give
\[
 \sum_{j=1}^J a_j C(|v_j-1/2|)\le\sum_{j=1}^J a_j b_j=\rho\le r .
\]
Thus $\boldsymbol v\in V_r$ and $|\theta-m|\le R(r)$. The maximum defining $R(r)$ exists by compactness of $D_r$. The same bound applies to arbitrary bin masses $(\boldsymbol p',\boldsymbol q')$, with equality at the vectors $v_j=1/2\pm d_j^*$ for a maximizing $\boldsymbol d^*$ at those masses.

For attainability, choose a maximizer $\boldsymbol d^*\in D_r$ at $(\boldsymbol p,\boldsymbol q)$ with $d_j^*=0$ on $\act^c$. For each $t\in[-1,1]$, Lemma~\ref{lem:A} supplies atomless conditional laws $\nu_{0j},\nu_{1j}$ on an open subinterval of $I_j$ with win probability $1/2+t d_j^*$ and cost $C(|t|d_j^*)$. On $\act^c$, take the two laws to be uniform on the same finite subinterval. Form the marginal laws
\[
 P_0=\sum_{j=1}^J p_j\nu_{0j},\qquad P_1=\sum_{j=1}^J q_j\nu_{1j}.
\]
They are atomless and have the prescribed bin masses. The total cost is at most $\sum_j a_j C(d_j^*)\le r$, and the AUC is $m+tR(r)$. Varying $t$ gives the whole interval. At $r\ge\sat$ the radius is $R_{\max}$, which recovers the unrestricted bound in Lemma~3 of Section~5.1 of \citet{Cormode2023}.
\end{proof}

\Needspace{4\baselineskip}\begin{proof}[\hypertarget{proof:thm:B1}{Proof} of Theorem~\ref{thm:B1}]

Claim~1 ($\act=\emptyset\implies R(r)=0$) and the boundary values $R(0)=0$ and $R(r)=R_{\max}$ for $r\in[\sat,1]$ when $\act\ne\emptyset$ follow directly from $C(0)=0$, $C(1/2)=1$, and $\sum_{j\in\act} c_j(1/2-d_j)\ge 0$ for all $\boldsymbol{d}\in D_r$.

At saturation, each positive $c_j$ forces $d_j=1/2$ for equality in the objective bound. These coordinates cost $\sat$, so the remaining coordinates must vanish at $r=\sat$. For $r>\sat$ they may use any remaining budget.

Now suppose $\act\ne\emptyset$ and $0<r<\sat$. Notice that $\Phi_B(0)=0$, $\lim_{h\to\infty}\Phi_B(h)=\sum_{j\in\act}a_j=\sat$, and $\Phi_B$ is continuous and strictly increasing on $[0,\infty)$ because $\beta$ is continuous and strictly increasing on $[0,\infty)$ and $z_j>0$ for $j\in\act$. Hence $\Phi_B(h)=r$ possesses a unique root $h>0$. Set $\lambda=1/(2h)>0$. Because $C$ is continuous and strictly convex with $C'(d)=6d$ on $[0,1/3]$ and $C'(d)=2/\sqrt{3-6d}$ on $(1/3,1/2)$, the single-coordinate Lagrangian $d\mapsto c_j d-\lambda a_j C(d)$ is strictly concave on $[0,1/2]$ and uniquely maximized at $d_j^*=G(hz_j)$ (which equals $0$ when $j\notin\act$, where $c_j=z_j=0$). For every $\boldsymbol{d}\in D_r$,
\[
 \sum_{j=1}^J c_j d_j
 \le \sum_{j=1}^J c_j d_j^* + \lambda\left(\sum_{j=1}^J a_j C(d_j)-r\right)
 \le \sum_{j=1}^J c_j d_j^*,
\]
The budget equation gives equality at $\boldsymbol d^*$, and strict concavity of each coordinate objective makes the first inequality strict for $\boldsymbol d\ne\boldsymbol d^*$.

Because $\beta'(t)=2t G'(t)>0$ for all $t>0$ (specifically $2t/3$ on $(0,1)$ and $2/(3t^2)$ on $(1,\infty)$), the chain rule yields
\[
 \Phi_R'(h)=\sum_{j\in\act} c_j z_j G'(hz_j)=\sum_{j\in\act} a_j z_j^2 G'(hz_j),\qquad
 \Phi_B'(h)=\sum_{j\in\act} a_j z_j\cdot 2(hz_j)G'(hz_j)=2h\Phi_R'(h)>0.
\]
Hence $R'(r)=\Phi_R'(h)/\Phi_B'(h)=1/(2h)$ decreases strictly in $r\in(0,\sat)$, proving that $R$ is strictly increasing and strictly concave on $[0,\sat]$. Continuity on $[0,1]$ follows from continuity at $r=0$ and $\lim_{h\to\infty}\Phi_R(h)=\sum_{j\in\act} c_j/2=R_{\max}$.

For Claim~3, $Z=\max_j z_j>0$ when $\act\ne\emptyset$. Notice $r_0=\Phi_B(1/Z)=S/(3Z^2)\le (\sum_{j\in\act} a_j)/3=\sat/3$. At $r=0$, feasibility forces $\boldsymbol{d}=0$ and the equation has the root $h=0$. For $0<r\le r_0$, the unique positive root satisfies $h\le 1/Z$, so $hz_j\le 1$ for all $j$, yielding $\Phi_B(h)=h^2 S/3=r$ and $R(r)=\Phi_R(h)=hS/3=\sqrt{rS/3}$. Since $C(d)>3d^2$ strictly for $d\in(1/3,1/2]$, weighted Cauchy--Schwarz gives for any $\boldsymbol{d}\in D_r$:
\[
 \Bigl(\sum_{j=1}^J c_j d_j\Bigr)^2
 =\Bigl(\sum_{j=1}^J a_j z_j d_j\Bigr)^2
 \le \Bigl(\sum_{j=1}^J a_j z_j^2\Bigr)\Bigl(\sum_{j=1}^J a_j d_j^2\Bigr)
 \le S\sum_{j=1}^J a_j\frac{C(d_j)}{3}
 \le \frac{rS}{3},
\]
with equality if and only if $d_j=k z_j\le 1/3$ for all $j$ and $\sum_{j=1}^J a_j C(d_j)=r$, which holds if and only if $0\le r\le r_0$. Thus $R(r)<\sqrt{rS/3}$ for all $r\in(r_0,1]$. \end{proof}

\begin{proof}[\hypertarget{proof:cor:C}{Proof} of Corollary~\ref{cor:C}]
Write $H_\tau=\{r\in[0,1]:L(r)\le\tau\}$. Theorem~\ref{thm:B1} gives
\[
 r^*(\tau)=
 \begin{cases}
 0,&\tau\ge m,\\
 +\infty,&\tau<L_{\min},\\
 \sat,&\tau=L_{\min}<m .
 \end{cases}
\]
Indeed, the first case has $0\in H_\tau$, the second has $H_\tau=\varnothing$, and the third has $H_\tau=[\sat,1]$. If $\act=\varnothing$, $L(r)=m=L_{\min}$ throughout and the first two cases suffice.

For $L_{\min}<\tau<m$, let $\delta=m-\tau$. The continuous, strictly increasing function $\Phi_R$ maps $[0,\infty)$ onto $[0,R_{\max})$. The unique $h>0$ with $\Phi_R(h)=\delta$ gives $H_\tau=[\Phi_B(h),1]$ by Theorem~\ref{thm:B1}. Hence $r^*(\tau)=\Phi_B(h)$. If $\delta\le S/(3Z)$, then $h\le1/Z$, so $\Phi_R(h)=hS/3$ and $\Phi_B(h)=h^2S/3$. Substitution yields $r^*(\tau)=3\delta^2/S$. By continuity of $L$, every finite first crossing belongs to $H_\tau$.
\end{proof}

\Needspace{4\baselineskip}
\begin{proof}[\hypertarget{proof:prop:kl}{Proof} of Proposition~\ref{prop:kl}]
For a fixed budget, apply Lemma~\ref{lem:D1} in Section~\ref{subsec:pointwise-kl-support} of Appendix~\ref{app:confidence-support} to fixed kernels attaining the population endpoints. For simultaneous coverage over $r\ge\rho$, use the true win vector as a single feasible kernel. For the threshold statement, apply the same one-sided bound to a fixed lower witness at $r^*(\tau)$ and invert $\ell_\delta$; Proposition~\ref{prop:D2} and Corollary~\ref{cor:D3} in Appendix~\ref{app:confidence-support} detail the coverage and boundary cases.
\end{proof}

\Needspace{4\baselineskip}
\begin{proof}[\hypertarget{proof:prop:band}{Proof} of Proposition~\ref{prop:band}]
For DKW, the simultaneous event in Proposition~\ref{prop:E3} and Lemma~\ref{lem:E2} in Appendix~\ref{app:confidence-support} give the corner bounds for every $r$. For the cutpoint Clopper--Pearson construction, Bonferroni gives simultaneous CDF coverage, and the same projection argument orders the feasible corners. Continuity over the connected CDF box gives the full interval between the corner endpoints. The DKW frontier guarantee follows by first-hitting inversion. Section~\ref{subsec:simultaneous-bands-support} of Appendix~\ref{app:confidence-support} gives the detailed endpoint argument. Corollaries~\ref{cor:E4} and~\ref{cor:E5} in Appendix~\ref{app:confidence-support} give the DKW inversions.
\end{proof}

\Needspace{6\baselineskip}
\begin{proof}[\hypertarget{proof:thm:level}{Proof} of Theorem~\ref{thm:level}]
Fix $\vartheta=(\boldsymbol p,\boldsymbol q)$ in the null and choose its optimal allocation $\boldsymbol d^*=\boldsymbol d^*(\boldsymbol p\odot\boldsymbol q;r)$ from Theorem~\ref{thm:B1}. Let $\mathbf K_\vartheta=\mathbf K(\boldsymbol d^*)$, and write
\[
 R=\mathbf K_\vartheta\boldsymbol q,\qquad
 S=\mathbf K_\vartheta^{\mathsf T}\boldsymbol p,\qquad
 L_\vartheta=\boldsymbol p^{\mathsf T}\mathbf K_\vartheta\boldsymbol q.
\]
Put $f_i=R_i-L_\vartheta$, $g_j=S_j-L_\vartheta$, and
$D_{ij}=K_{\vartheta,ij}-R_i-S_j+L_\vartheta$. The row and column centering identities give
$\boldsymbol p^{\mathsf T}f=\boldsymbol q^{\mathsf T}g=0$ and
$D\boldsymbol q=D^{\mathsf T}\boldsymbol p=0$. For independent samples with observed bin indices $B(X_s)$ and $B(Y_t)$, the centered witness average has the decomposition
\begin{equation}\label{eq:projection-decomposition}
 \frac{1}{n_0n_1}\sum_{s,t}K_{\vartheta,B(X_s),B(Y_t)}-L_\vartheta
 =
 \underbrace{\frac1{n_0}\sum_s f_{B(X_s)}
              +\frac1{n_1}\sum_t g_{B(Y_t)}}_{P_\vartheta}
 +\underbrace{\frac1{n_0n_1}\sum_{s,t}D_{B(X_s),B(Y_t)}}_{W_\vartheta}.
\end{equation}
Combining \eqref{eq:projection-decomposition} with sample minimization gives
\begin{equation}\label{eq:fixed-witness-supply}
 \widehat L-L_\vartheta
 \le \widehat{\boldsymbol p}^{\mathsf T}\mathbf K_\vartheta\widehat{\boldsymbol q}-L_\vartheta
 =P_\vartheta+W_\vartheta .
\end{equation}
The increments $Z_\ell$ in $P_\vartheta$ are independent and centered. Their variance, third-moment sum, and fourth-moment sum are
\[
 V=\sum_\ell \E Z_\ell^2
 =\frac{\boldsymbol p^{\mathsf T}f^{\odot2}}{n_0}
  +\frac{\boldsymbol q^{\mathsf T}g^{\odot2}}{n_1},\qquad
 \kappa=\sum_\ell \E Z_\ell^3
 =\frac{\boldsymbol p^{\mathsf T}f^{\odot3}}{n_0^2}
  +\frac{\boldsymbol q^{\mathsf T}g^{\odot3}}{n_1^2},
\]
\[
 m_4=\sum_\ell \E Z_\ell^4
 =\frac{\boldsymbol p^{\mathsf T}f^{\odot4}}{n_0^3}
  +\frac{\boldsymbol q^{\mathsf T}g^{\odot4}}{n_1^3},
 \qquad b=\max_\ell\operatorname{ess\,sup}|Z_\ell| .
\]
Here $m_4$ is the sum of fourth moments of the increments. The fourth moment of $P_\vartheta$ also contains the cross terms
$6\sum_{\ell<k}\E Z_\ell^2\E Z_k^2$.

Each of the $2J$ marginal bin counts is binomial. Invert both inclusive tails at $\gamma/(4J)$, using 0 and 1 at the endpoints. Monotonicity in the binomial parameter and the inclusive probability transform bound each endpoint's failure probability by $\gamma/(4J)$. Bonferroni therefore gives
\[
 \Pr_\vartheta(\vartheta\notin\mathcal C)\le\gamma,
\]
including after outward rounding. The simultaneous-band construction described in Appendix~\ref{app:confidence-support}, Section~\ref{subsec:simultaneous-bands-support} instead applies binomial intervals to cumulative counts at the $J-1$ internal cutpoints, with tail allocation $\alpha/[4(J-1)]$.

The double-centered remainder has mean zero and
\[
 v_D=\operatorname{Var}(W_\vartheta)=\frac{\E D_{B(X),B(Y)}^2}{n_0n_1}.
\]
Because $D$ has zero conditional means in both coordinates, terms in $W_\vartheta$ from disjoint pairs are independent, and terms sharing one observation have zero covariance after conditioning on that observation. The same centering gives $\operatorname{Cov}(W_\vartheta,P_\vartheta)=0$. Appendix~\ref{app:probability-boxes} gives uniform box bounds for $v_D$ and the projection moments.

Set $e_\vartheta=\sqrt{v_D(1-\delta_D)/\delta_D}$. If $v_D>0$, the one-sided second-moment inequality applied to the shifted remainder with $c=v_D/e_\vartheta$ yields
\[
 \Pr_\vartheta(W_\vartheta>e_\vartheta)
 \le \frac{v_D+c^2}{(e_\vartheta+c)^2}=\delta_D.
\]
If $v_D=0$, then $W_\vartheta=0$ almost surely.

For every probability box $B\subseteq\mathcal C$, the outward enclosures in Appendices~\ref{app:probability-boxes}--\ref{app:numerical-cert} give $e_B\ge e_\vartheta$ for $\vartheta\in B$ and a lower threshold $x_B\le\widehat L-\tau-e_B$. Appendix~\ref{app:tail-bounds} proves that the certified bound at $x_B$ is an upper bound on the inclusive projection tail, with Bennett's inequality available whenever the Fourier conditions cannot be verified.

The queued and removed boxes cover the root throughout the search because closed bisection preserves their union. A box is removed for infeasibility, for $L^->\tau$, or after its tail bound at $x_B$ is at most $\delta_P$. The algorithm returns non-rejection at an unresolved center or at a resource limit. If it rejects while $\vartheta\in\mathcal C$ under the null, a box containing $\vartheta$ must therefore have met the tail rule. The lower-endpoint certificate cannot remove such a box because $L^-\le L_\vartheta\le\tau$.

Let $s_\vartheta(t)=\Pr_\vartheta(P_\vartheta\ge t)$. On the additional event $W_\vartheta\le e_\vartheta$, equation~\eqref{eq:fixed-witness-supply} and $L_\vartheta\le\tau$ imply
\[
 x_B\le\widehat L-\tau-e_\vartheta\le P_\vartheta,
 \qquad
 s_\vartheta(P_\vartheta)\le s_\vartheta(x_B)\le\delta_P.
\]
The event $\{s_\vartheta(P_\vartheta)\le\delta_P\}$ is empty or consists of values at or above the first support point whose inclusive tail is at most $\delta_P$. Its probability is at most $\delta_P$. Thus
\begin{equation}\label{eq:rejection-events}
 \{\mathrm{reject}\}\subseteq
 \{\vartheta\notin\mathcal C\}\cup\{W_\vartheta>e_\vartheta\}
 \cup\{s_\vartheta(P_\vartheta)\le\delta_P\}.
\end{equation}
Applying the union bound to \eqref{eq:rejection-events} gives $\gamma+\delta_D+\delta_P=0.025$ for every null parameter at the prespecified pair.
\end{proof}

\FloatBarrier
\section{Auxiliary results for confidence intervals and simultaneous bands}\label{app:confidence-support}
\subsection{Pointwise KL inference}\label{subsec:pointwise-kl-support}
Write $D(x\Vert y)=x\log(x/y)+(1-x)\log((1-x)/(1-y))$ for $x\in[0,1]$ and $y\in(0,1)$, extended by continuity at $x\in\{0,1\}$ and $+\infty$ at incompatible boundaries. For $\delta\in(0,1)$ and $k=\min(n_0,n_1)\ge 1$, define
\[
 \ell_\delta(x)=\inf\{y\in[0,x]:k\,D(x\Vert y)\le\log(1/\delta)\},\qquad
 u_\delta(x)=1-\ell_\delta(1-x),
\]
so $\ell_\delta(0)=0$, $\ell_\delta(1)=\delta^{1/k}$, $u_\delta(0)=1-\delta^{1/k}$, $u_\delta(1)=1$, and both $\ell_\delta,u_\delta$ are continuous and strictly increasing on $[0,1]$.

\begin{lemma}\label{lem:D1}
Let $h:\mathbb R^2\to[0,1]$ be a fixed measurable kernel. Put $\mu=\E h(X_1,Y_1)$ and
\[
 W=\frac1{n_0 n_1}\sum_{s=1}^{n_0}\sum_{t=1}^{n_1} h(X_s,Y_t).
\]
Then $\Prob(W\ge x)\le e^{-kD(x\Vert\mu)}$ for $x\ge\mu$, and $\Prob(W\le x)\le e^{-kD(x\Vert\mu)}$ for $x\le\mu$. In particular,
\[
 \Prob\{\ell_\delta(W)>\mu\}\le\delta,\qquad
 \Prob\{u_\delta(W)<\mu\}\le\delta.
\]
\end{lemma}

\begin{proof}
For independent uniform permutations $\pi,\sigma$ independent of $(X,Y)$, put
\[
V=k^{-1}\sum_{s=1}^k h(X_{\pi(s)},Y_{\sigma(s)}).
\]
This is an average of $k$ iid $[0,1]$-valued variables with $\E(V\mid X,Y)=W$. Conditional Jensen and $e^{tz}\le 1-z+ze^t$ on $z\in[0,1]$ give
\[
\E e^{tkW}\le \E e^{tkV}\le(1-\mu+\mu e^t)^k,\qquad t\in\mathbb R,
\]
as in Theorem~1 and Section~5b of \citet{Hoeffding1963}. Optimizing Markov's inequality over $t\ge 0$ (or $t\le 0$) yields $e^{-kD(x\Vert\mu)}$. Monotonicity of $x\mapsto D(x\Vert\mu)$ on $[\mu,1]$ implies $\{\ell_\delta(W)>\mu\}\subseteq\{W>\mu,\,kD(W\Vert\mu)>\log(1/\delta)\}$, of probability at most $\delta$; applying this to $1-h$ gives the upper bound.
\end{proof}

\begin{proposition}\label{prop:D2}
Fix $\alpha\in(0,1)$, put $\delta=\alpha/2$, and define $I_{\KL}(r)=[\ell_\delta(\hat L(r)),u_\delta(\hat U(r))]$ using the empirical endpoints $\hat L(r)=L(\hat{\boldsymbol{p}},\hat{\boldsymbol{q}},\boldsymbol{a};r)$ and $\hat U(r)=U(\hat{\boldsymbol{p}},\hat{\boldsymbol{q}},\boldsymbol{a};r)$. For each prespecified $r\in[0,1]$,
\[
 \Prob\bigl([L(r),U(r)]\subseteq I_{\KL}(r)\bigr)\ge 1-\alpha.
\]
\end{proposition}

\begin{proof}
Let $\boldsymbol v^-$ and $\boldsymbol v^+$ attain $L(r)$ and $U(r)$, respectively. Both belong to $V_r$ for every realization of the counts. Hence
\[
 \hat L(r)\le T(\hat{\boldsymbol p},\hat{\boldsymbol q};\boldsymbol v^-),
 \qquad
 T(\hat{\boldsymbol p},\hat{\boldsymbol q};\boldsymbol v^+)\le \hat U(r).
\]
Apply Lemma~\ref{lem:D1} to these fixed kernels and use monotonicity of $\ell_\delta$ and $u_\delta$. Each endpoint has failure probability at most $\delta$, so the probability that either endpoint fails is at most $\alpha$ \citep{Imbens2004}.
\end{proof}

\begin{corollary}\label{cor:D3}
\begin{enumerate}
\item Under (M1)--(M4), let $(P_0,P_1)$ have true AUC $\theta$ and true Brier information $\rho$. Then $\Prob(\forall r\in[\rho,1],\,\theta\in I_{\KL}(r))\ge 1-\alpha$.
\item Fix a single prespecified threshold $\tau\in\mathbb{R}$ and error level $\delta\in(0,1)$. Define $\underline r_{\KL,\delta}(\tau)=\inf\{r\in[0,1]:\ell_\delta(\hat L(r))\le\tau\}$ ($\inf\emptyset=+\infty$). Then $\Prob(\underline r_{\KL,\delta}(\tau)\le r^*(\tau))\ge 1-\delta$, and $\underline r_{\KL,\delta}(\tau)=r^*(\hat{\boldsymbol{p}},\hat{\boldsymbol{q}},\boldsymbol{a};x_\delta(\tau))$, where $x_\delta(\tau)=-\infty$ for $\tau<0$, $0$ for $\tau=0$, $1$ for $\tau\ge\delta^{1/k}$, and the unique root in $(\tau,1)$ of $k\,D(x\Vert\tau)=\log(1/\delta)$ for $0<\tau<\delta^{1/k}$.
\end{enumerate}
\end{corollary}

\begin{proof}
For part~1, the true win vector $\boldsymbol v^0$ belongs to $V_\rho$ and thus to $V_r$ for every $r\ge\rho$. Write $W_0=T(\hat{\boldsymbol p},\hat{\boldsymbol q};\boldsymbol v^0)$. Its population value is $\theta$, and $\hat L(r)\le W_0\le\hat U(r)$ for all $r\ge\rho$. A single paired KL event for $W_0$ gives coverage simultaneously over these budgets.

For part~2, the claim holds immediately if $r^*(\tau)=+\infty$. Otherwise write $r^*=r^*(\tau)$ and choose the fixed lower witness at $r^*$. Lemma~\ref{lem:D1} and Corollary~\ref{cor:C} in Appendix~\ref{app:proofs} give
\[
 \ell_\delta(\hat L(r^*))\le L(r^*)\le\tau
\]
with probability at least $1-\delta$. Thus $r^*$ belongs to the set defining $\underline r_{\KL,\delta}(\tau)$ on this event. On $[0,1]$, $\ell_\delta$ is continuous and strictly increasing with range $[0,\delta^{1/k}]$. Inverting it gives $\ell_\delta(\hat L(r))\le\tau$ exactly when $\hat L(r)\le x_\delta(\tau)$, including the endpoint cases in the statement. Taking the infimum yields the formula.
\end{proof}

\subsection{Simultaneous confidence bands}\label{subsec:simultaneous-bands-support}

For $g\in\{0,1\}$ and $j\in\{0,1,\ldots,J\}$, define the population and empirical bin CDFs
\[
 F_0(j)=\sum_{\ell=1}^j p_\ell,\quad \hat F_0(j)=\sum_{\ell=1}^j\hat p_\ell,\qquad
 F_1(j)=\sum_{\ell=1}^j q_\ell,\quad \hat F_1(j)=\sum_{\ell=1}^j\hat q_\ell,
\]
with $F_g(0)=\hat F_g(0)=0$ and $F_g(J)=\hat F_g(J)=1$ (empty maxima for $J=1$ equal $0$).

\begin{lemma}\label{lem:E1}
Fix $\alpha\in(0,1)$ and set
\[
 \varepsilon_0=\sqrt{\frac{\log(4/\alpha)}{2n_0}},\qquad
 \varepsilon_1=\sqrt{\frac{\log(4/\alpha)}{2n_1}}.
\]
Define
\[
 \mathcal{E}_g=\left\{\max_{1\le j\le J-1}|\hat F_g(j)-F_g(j)|\le\varepsilon_g\right\}\;(g=0,1),\qquad
 \mathcal{E}=\mathcal{E}_0\cap\mathcal{E}_1,
\]
\[
 \mathcal{E}_-=\left\{\max_{1\le j\le J-1}(\hat F_0(j)-F_0(j))\le\varepsilon_0\right\}
 \cap\left\{\max_{1\le j\le J-1}(F_1(j)-\hat F_1(j))\le\varepsilon_1\right\}.
\]
For all $J\ge 1$, $n_0,n_1\ge 1$, and $\boldsymbol{p},\boldsymbol{q}\in\Delta_J$, one has $\mathcal{E}\subseteq\mathcal{E}_-$, $\Prob(\mathcal{E})\ge(1-\alpha/2)^2>1-\alpha$, and $\Prob(\mathcal{E}_-)\ge(1-\alpha/4)^2>1-\alpha/2$.
\end{lemma}

\begin{proof}
For $J=1$, both events are certain by the empty-maximum convention. For $J\ge2$, generate the group-$g$ bin indices from independent uniforms $\xi_{g,s}$ by
\[
 Z_{g,s}=\min\{j\in\{1,\ldots,J\}:\xi_{g,s}\le F_g(j)\}.
\]
Then $\{Z_{g,s}\le j\}=\{\xi_{g,s}\le F_g(j)\}$, including at flat steps of $F_g$. Thus the generated indices have the sampling law of the observed bins. If $\hat G_g$ is the empirical CDF of the uniforms, the generated empirical bin CDF is $\hat G_g(F_g(j))$.

The maximum deviation over bin cutpoints is at most the supremum deviation of $\hat G_g$ over $[0,1]$. The same comparison applies separately to each signed deviation. At $\varepsilon_g$, the two-sided Dvoretzky--Kiefer--Wolfowitz--Massart bound is $2e^{-2n_g\varepsilon_g^2}=\alpha/2$, and the one-sided bound is $e^{-2n_g\varepsilon_g^2}=\alpha/4$ \cite{Massart1990,LeBoudec2026}. Independence of the two groups gives
\[
 \Prob(\mathcal E)\ge(1-\alpha/2)^2,\qquad
 \Prob(\mathcal E_-)\ge(1-\alpha/4)^2.
\]
The inclusion $\mathcal E\subseteq\mathcal E_-$ follows from the definitions.
\end{proof}

\begin{lemma}\label{lem:E2}
For any $\boldsymbol{p}',\tilde{\boldsymbol{p}},\boldsymbol{q}',\tilde{\boldsymbol{q}}\in\Delta_J$ with cumulative sums $F_{\boldsymbol{p}'},F_{\tilde{\boldsymbol{p}}},F_{\boldsymbol{q}'},F_{\tilde{\boldsymbol{q}}}$ and any $\boldsymbol{v}\in[0,1]^J$,
\begin{align}
 T(\boldsymbol{p}',\boldsymbol{q}';\boldsymbol{v})-T(\tilde{\boldsymbol{p}},\tilde{\boldsymbol{q}};\boldsymbol{v})
 &=\sum_{i=1}^{J-1}\bigl(F_{\boldsymbol{p}'}(i)-F_{\tilde{\boldsymbol{p}}}(i)\bigr)\bigl[q'_i v_i+q'_{i+1}(1-v_{i+1})\bigr] \notag\\
 &\quad+\sum_{j=1}^{J-1}\bigl(F_{\tilde{\boldsymbol{q}}}(j)-F_{\boldsymbol{q}'}(j)\bigr)\bigl[\tilde p_j(1-v_j)+\tilde p_{j+1}v_{j+1}\bigr]. \label{eq:sbp}
\end{align}
In particular, if $F_{\tilde{\boldsymbol{p}}}(j)\le F_{\boldsymbol{p}'}(j)\le F_{\bar{\boldsymbol{p}}}(j)$ and $F_{\bar{\boldsymbol{q}}}(j)\le F_{\boldsymbol{q}'}(j)\le F_{\tilde{\boldsymbol{q}}}(j)$ for all $1\le j\le J-1$, then for every fixed positive weight vector $\boldsymbol{a}\in\Delta_J$ and every $r\in[0,1]$,
\[
 L(\tilde{\boldsymbol{p}},\tilde{\boldsymbol{q}},\boldsymbol{a};r)\le L(\boldsymbol{p}',\boldsymbol{q}',\boldsymbol{a};r)\le U(\boldsymbol{p}',\boldsymbol{q}',\boldsymbol{a};r)\le U(\bar{\boldsymbol{p}},\bar{\boldsymbol{q}},\boldsymbol{a};r).
\]
\end{lemma}

\begin{proof}
For $J=1$, both sums in \eqref{eq:sbp} are empty. For $J\ge2$, fix $\boldsymbol v$ and write $T(\boldsymbol p,\boldsymbol q)=T(\boldsymbol p,\boldsymbol q;\boldsymbol v)$. Define the row and column means
\[
 g_i=\sum_{j=1}^J q'_j\mathbf K_{\boldsymbol v}(i,j),
 \qquad h_j=\sum_{i=1}^J\tilde p_i\mathbf K_{\boldsymbol v}(i,j).
\]
When a row or column index increases by one, the kernel changes only at the two adjacent diagonal positions. Its definition therefore gives
\begin{equation}\label{eq:kernel-adjacent}
\begin{aligned}
 g_i-g_{i+1}&=q'_i v_i+q'_{i+1}(1-v_{i+1})\ge0,\\
 h_{j+1}-h_j&=\tilde p_j(1-v_j)+\tilde p_{j+1}v_{j+1}\ge0.
\end{aligned}
\end{equation}

Change the first marginal while holding the second fixed, then change the second marginal. Summation by parts gives
\begin{equation}\label{eq:sbp-marginals}
\begin{aligned}
 T(\boldsymbol p',\boldsymbol q')-T(\tilde{\boldsymbol p},\boldsymbol q')
 &=\sum_{i=1}^{J-1}\{F_{\boldsymbol p'}(i)-F_{\tilde{\boldsymbol p}}(i)\}(g_i-g_{i+1}),\\
 T(\tilde{\boldsymbol p},\boldsymbol q')-T(\tilde{\boldsymbol p},\tilde{\boldsymbol q})
 &=\sum_{j=1}^{J-1}\{F_{\tilde{\boldsymbol q}}(j)-F_{\boldsymbol q'}(j)\}(h_{j+1}-h_j).
\end{aligned}
\end{equation}
The CDF differences vanish at 0 and $J$, so there are no boundary terms. Add the two equalities in \eqref{eq:sbp-marginals} and substitute \eqref{eq:kernel-adjacent} to obtain \eqref{eq:sbp}.

Under the stated CDF order, both sums are nonnegative. Applying the same argument to the upper corner gives
\[
 T(\tilde{\boldsymbol p},\tilde{\boldsymbol q})
 \le T(\boldsymbol p',\boldsymbol q')
 \le T(\bar{\boldsymbol p},\bar{\boldsymbol q}).
\]
These inequalities hold for every $\boldsymbol v\in V_r$. By Theorem~\ref{thm:B} in Appendix~\ref{app:proofs}, the same set $V_r$ applies to all three pairs of marginals. Taking minima for the lower endpoints and maxima for the upper endpoints gives the asserted bounds.
\end{proof}

Define the clipped CDF envelopes ($g\in\{0,1\}$) by $F_g^\pm(0)=0$, $F_g^\pm(J)=1$, and
\[
 F_g^-(j)=\max\bigl(0,\hat F_g(j)-\varepsilon_g\bigr),\qquad
 F_g^+(j)=\min\bigl(1,\hat F_g(j)+\varepsilon_g\bigr),\qquad j=1,\ldots,J-1,
\]
with differenced mass vectors $p_j^\pm=F_0^\pm(j)-F_0^\pm(j-1)$ and $q_j^\pm=F_1^\pm(j)-F_1^\pm(j-1)$ ($j=1,\ldots,J$). Set
\[
 L_{\DKW}(r)=L(\boldsymbol{p}^-,\boldsymbol{q}^+,\boldsymbol{a};r),\qquad U_{\DKW}(r)=U(\boldsymbol{p}^+,\boldsymbol{q}^-,\boldsymbol{a};r),\qquad r\in[0,1].
\]

\begin{proposition}\label{prop:E3}
Under (M1)--(M4) and $\alpha\in(0,1)$:
\begin{enumerate}
\item On every sample, $\boldsymbol{p}^\pm,\boldsymbol{q}^\pm\in\Delta_J$ and $0\le L_{\DKW}(r)\le\hat L(r)\le\hat U(r)\le U_{\DKW}(r)\le 1$ for all $r\in[0,1]$.
\item Let $\mathcal{B}=\{(\tilde{\boldsymbol{p}},\tilde{\boldsymbol{q}})\in\Delta_J\times\Delta_J:F_0^-\le F_{\tilde{\boldsymbol{p}}}\le F_0^+,\,F_1^-\le F_{\tilde{\boldsymbol{q}}}\le F_1^+\}$. For every $r\in[0,1]$,
\[
 [L_{\DKW}(r),U_{\DKW}(r)]=\bigcup_{(\tilde{\boldsymbol{p}},\tilde{\boldsymbol{q}})\in\mathcal{B}}[L(\tilde{\boldsymbol{p}},\tilde{\boldsymbol{q}},\boldsymbol{a};r),U(\tilde{\boldsymbol{p}},\tilde{\boldsymbol{q}},\boldsymbol{a};r)].
\]
\item On the single event $\mathcal{E}$ of Lemma~\ref{lem:E1}, $[L(r),U(r)]\subseteq[L_{\DKW}(r),U_{\DKW}(r)]$ simultaneously for all $r\in[0,1]$, so
\[
 \Prob\Bigl(\forall r\in[0,1],\;[L(r),U(r)]\subseteq[L_{\DKW}(r),U_{\DKW}(r)]\Bigr)\ge(1-\alpha/2)^2>1-\alpha.
\]
\end{enumerate}
\end{proposition}

\begin{proof}
Clipping $\hat F_g\pm\varepsilon_g$ is monotone and preserves the assigned endpoints at 0 and $J$. Thus $\boldsymbol p^\pm,\boldsymbol q^\pm$ lie in $\Delta_J$. Since $F_g^-\le\hat F_g\le F_g^+$, Lemma~\ref{lem:E2} gives the sample inclusions in part~1.

For part~2, both corners $(\boldsymbol p^-,\boldsymbol q^+)$ and $(\boldsymbol p^+,\boldsymbol q^-)$ belong to $\mathcal B$. The same lemma bounds the union by the two corner endpoints. The line segment joining the corners lies in $\mathcal B$. Along it, continuity of $m$ covers every value between the two corner mid-ranks. Each such value belongs to its own identified interval, and the corner intervals cover the remaining values up to the endpoints.

On $\mathcal E$, the two true CDFs lie between their clipped bounds, so $(\boldsymbol p,\boldsymbol q)\in\mathcal B$. Part~2 then gives the inclusion for every $r$ on this single event. All events are measurable on the finite count space. The probability bound follows from Lemma~\ref{lem:E1}.
\end{proof}

\begin{corollary}\label{cor:E4}
Define $\underline r_{\DKW}(\tau)=\inf\{r\in[0,1]:L_{\DKW}(r)\le\tau\}\in[0,1]\cup\{+\infty\}$ ($\inf\emptyset=+\infty$).
\begin{enumerate}
\item For every $\tau\in\mathbb{R}$, $\underline r_{\DKW}(\tau)=r^*(\boldsymbol{p}^-,\boldsymbol{q}^+,\boldsymbol{a};\tau)=\min_{(\tilde{\boldsymbol{p}},\tilde{\boldsymbol{q}})\in\mathcal{B}}r^*(\tilde{\boldsymbol{p}},\tilde{\boldsymbol{q}},\boldsymbol{a};\tau)$, evaluated directly by Corollary~\ref{cor:C} in Appendix~\ref{app:proofs} at $(\boldsymbol{p}^-,\boldsymbol{q}^+)$.
\item On every sample, $\{\forall r\in[0,1],\,L_{\DKW}(r)\le L(r)\}=\{\forall\tau\in\mathbb{R},\,\underline r_{\DKW}(\tau)\le r^*(\tau)\}=\{\forall\tau\in[0,1],\,\underline r_{\DKW}(\tau)\le r^*(\tau)\}$.
\item On the same event $\mathcal{E}$ of Proposition~\ref{prop:E3}, both $[L(r),U(r)]\subseteq[L_{\DKW}(r),U_{\DKW}(r)]$ for all $r\in[0,1]$ and $\underline r_{\DKW}(\tau)\le r^*(\tau)$ for all $\tau\in\mathbb{R}$ hold simultaneously with probability $>1-\alpha$; marginally, $\Prob(\forall\tau\in\mathbb{R},\,\underline r_{\DKW}(\tau)\le r^*(\tau))\ge\Prob(\mathcal{E}_-)>1-\alpha/2$.
\end{enumerate}
\end{corollary}

\begin{proof}
Part~1 follows from $L_{\DKW}(r)=L(\boldsymbol{p}^-,\boldsymbol{q}^+,\boldsymbol{a};r)$ and Proposition~\ref{prop:E3}(2). In part~2, if $L_{\DKW}(r)\le L(r)$ for all $r\in[0,1]$, then $\{r:L(r)\le\tau\}\subseteq\{r:L_{\DKW}(r)\le\tau\}$ for all $\tau\in\mathbb{R}$, so $\underline r_{\DKW}(\tau)\le r^*(\tau)$. Conversely, if $L_{\DKW}(r_0)>L(r_0)$ for some $r_0\in[0,1]$, put $\tau_0=L(r_0)\in[0,1]$, so $r^*(\tau_0)\le r_0$. Because $L_{\DKW}$ is continuous and nonincreasing on $[0,1]$ by Theorem~\ref{thm:B1} in Appendix~\ref{app:proofs}, either $L_{\DKW}(1)>\tau_0$ (giving $\underline r_{\DKW}(\tau_0)=+\infty>r_0$) or $r_1=\underline r_{\DKW}(\tau_0)\in[0,1]$ satisfies $L_{\DKW}(r_1)\le\tau_0<L_{\DKW}(r_0)$, forcing $r_1>r_0\ge r^*(\tau_0)$. Part~3 follows immediately from Lemma~\ref{lem:E1}, Proposition~\ref{prop:E3}(3), and part~2.
\end{proof}

\begin{corollary}\label{cor:E5}
For a standalone one-sided simultaneous lower confidence bound at level $1-\delta$ ($\delta\in(0,1)$) without an accompanying upper confidence band, replacing $\varepsilon_g$ by $\varepsilon_{g,\mathrm{1s}}(\delta)=\sqrt{\log(2/\delta)/(2n_g)}$ yields $\Prob(\forall\tau\in\mathbb{R},\,\underline r_{\DKW,\mathrm{1s}}(\tau;\delta)\le r^*(\tau))=\Prob(\forall r\in[0,1],\,L_{\DKW,\mathrm{1s}}(r;\delta)\le L(r))\ge(1-\delta/2)^2>1-\delta$, with $\varepsilon_{g,\mathrm{1s}}(\alpha/2)=\varepsilon_g(\alpha)$.
\end{corollary}

\begin{proof}
The inversion in Corollary~\ref{cor:E4} applies to the one-sided band. Its calibration follows by applying the one-sided uniform bound to each of the two oriented CDF deviations with tail $\delta/2$. Independence yields the stated product bound, and the same summation-by-parts and first-hitting argument gives the one-sided frontier event.
\end{proof}

For the Clopper--Pearson cutpoint band, the CDF endpoints are known when $J=1$. When $J>1$, each cumulative count at an internal cut $j$ is binomial with parameter $F_g(j)$. Invert both inclusive binomial tails at $\alpha/[4(J-1)]$ for each group and cut. Bonferroni's inequality then gives simultaneous coverage at least $1-\alpha$ for the $2(J-1)$ internal CDF values.

Project each lower sequence to its prefix maxima and each upper sequence to its suffix minima. Since the true CDFs are nondecreasing, this projection preserves coverage and gives valid ordered CDF bounds after adding 0 and 1 at the endpoints. Differencing the ordered bounds gives four valid mass vectors. Lemma~\ref{lem:E2} then projects the sharp identification endpoints to the corresponding opposite group corners at every budget. The endpoint values are attained within the CDF box, and continuity along line segments in the connected product of the CDF boxes and feasible win-vector set covers every value between them.

Both procedures use ordered-CDF projection. DKW inversion gives simultaneous lower limits for $r^*(\tau)$, with $+\infty$ for an empty hitting set; the fixed-pair test instead uses single-bin Clopper--Pearson intervals at tail level $\gamma/(4J)$. Appendix~\ref{app:proofs} uses these intervals in the proof of Theorem~\ref{thm:level}.

\FloatBarrier
\section{Uniform bounds over probability boxes}\label{app:probability-boxes}

Appendix~\ref{app:proofs} establishes the optimal allocation. Appendix~\ref{app:probability-boxes} records the extended conventions and outward bounds needed to enclose that allocation, the fixed-witness projection moments, and the degenerate remainder over a simplex box.

\subsection{Allocation envelopes over collision boxes}

Let $\boldsymbol{p},\boldsymbol{q}$ be the two bin probability vectors, let $c_j=p_jq_j$, and fix positive weights $a_j$ with $\sum_j a_j=1$ and budget $r\in[0,1]$. The allocation argument permits any nonnegative collision vector $c$, including the corners of an enclosing box. The single-bin cost and its inverse marginal rule give
\[
\begin{aligned}
 C(d)&=\begin{cases}3d^2,&0\le d\le1/3,\\
             1-\frac23\sqrt{3-6d},&1/3<d\le1/2,\end{cases}\\
 G(t)&=\begin{cases}t/3,&0\le t\le1,\\
             1/2-1/(6t^2),&t>1,\end{cases}\quad
 \beta(t)=C(G(t))=\begin{cases}t^2/3,&0\le t\le1,\\
             1-2/(3t),&t>1.\end{cases}
\end{aligned}
\]
Write $\mathcal D_r=\{\boldsymbol{d}\in[0,1/2]^J:\sum_j a_jC(d_j)\le r\}$, $A(c)=\sum_{c_j>0}a_j$, and $B_c(h)=\sum_j a_j\beta(hc_j/a_j)$ for finite $h\ge0$. Since $\beta$ is continuous and strictly increasing to $1$, $B_c$ is continuous and strictly increasing from $0$ to $A(c)$ for any $c$ with at least one active coordinate. Its finite range excludes $A(c)$. For $c=0$ the budget function is identically zero.

Set $h(c;0)=0$. For $r>0$ set $h(c;r)=\inf\{h\ge0:B_c(h)\ge r\}$, with an empty infimum equal to $+\infty$. Thus $0<r<A(c)$ has a unique finite positive root and $r\ge A(c)$ has infinite scale. Use the extended function
\[
 g_j(h,s)=\begin{cases}
 0,&h=0\text{ or } s=0,\\
 1/2,&h=+\infty,\ s>0,\\
 G(hs/a_j),&0<h<+\infty,\ s>0.
 \end{cases}
\]
It is nondecreasing in each argument. In the interior $h$ is the reciprocal of twice the budget multiplier $\lambda$.

The optimality and interior formula for the allocation are proved in Theorem~\ref{thm:B1} of Appendix~\ref{app:proofs}. Here we record the extended boundary convention and the monotonicity properties needed for interval certification.
When $c\le c'$ coordinatewise, $B_c(h)\le B_{c'}(h)$ for each finite $h$, since $\beta$ is increasing and the weights are positive. The inverse sets defining $h$ are nested, so $h(c';r)\le h(c;r)$, including the empty-set case and changes in active support. The two zero-budget scales are both zero. No assumption about continuity of the active set is used.

Any feasible coordinate satisfies $a_j C(d_j)\le r$, since all other costs are nonnegative. For $s=r/a_j$ the exact inverse cap is
\[
 H_j(r)=
 \begin{cases}
 \sqrt{s/3},&0\le s\le1/3,\\
 1/2-\frac38(1-s)^2,&1/3<s<1,\\
 1/2,&s\ge1.
 \end{cases}
\]
Both junctions agree. The middle expression follows from $\sqrt{3-6d}=(3/2)(1-s)$. Consequently if $0\le c^-\le c\le c^+$,
\begin{equation}\label{eq:allocation-envelope}
 g_j(h(c^+;r),c_j^-)
 \ \le\ d_j^*(c;r)\ \le\
 \min\{g_j(h(c^-;r),c_j^+),H_j(r)\}.
\end{equation}
The lower and upper inequalities first compare the inverse scales and then the corresponding coordinate inputs, using monotonicity of $g_j$ in both arguments. The hard cap gives the second upper bound. When $c_j^+=0$ both endpoints vanish. If $c_j^-=0<c_j^+$, the lower endpoint is zero and even an infinite upper scale retains the single-bin cap. If $r\ge A(c^+)>0$, every candidate is saturated. The formula uses $1/2$ at definitely active coordinates and allows zero at potentially inactive coordinates. The all-zero support, $r=0$, one bin, unequal fixed weights, both cost pieces, and equality at a saturation threshold require no additional cases. The enclosing vectors may use different corners and need not be feasible allocations themselves.

The implementation uses a verified lower scale for $c^+$ and upper scale for $c^-$, with directed evaluations of the gain and the cap. Monotonicity gives the same one-sided implications for these machine outputs.

Let $P=\{\boldsymbol{p}:\ell_P\le \boldsymbol{p}\le u_P,\ \mathbf{1}^{\mathsf T}\boldsymbol{p}=1\}$ and $Q$ be the corresponding group-1 box, with bounds in $[0,1]$. A simplex box is nonempty exactly when each lower bound is at most its upper bound and $\sum\ell_j\le1\le\sum u_j$. For sufficiency, add $1-\sum\ell_j$ to the lower vector in any order, respecting each capacity $u_j-\ell_j$. Tightening using $1-\sum_{k\ne j}u_k$ and $1-\sum_{k\ne j}\ell_k$ preserves precisely the same feasible points.

For these boxes, define $c_j^-=\ell_{P,j}\ell_{Q,j}$ and $c_j^+=u_{P,j}u_{Q,j}$. Then $c_j^-\le p_jq_j\le c_j^+$. Let $d_j^L$ and $d_j^U$ denote the lower and upper bounds in \eqref{eq:allocation-envelope}.

Write $\operatorname{lo}_P(w)$ and $\operatorname{hi}_P(w)$ for extrema of $w^{\mathsf T}\boldsymbol{p}$. To maximize, fill in decreasing order of $w_j$. Indeed, for any real $\lambda$ and any feasible $\boldsymbol{p}$,
\[
 w^{\mathsf T}\boldsymbol{p}\le\lambda+\sum_{w_j>\lambda}(w_j-\lambda)u_j+
                            \sum_{w_j<\lambda}(w_j-\lambda)\ell_j .
\]
At the first coefficient $\lambda$ where cumulative capacity accommodates the remaining mass, the filling vector attains equality. Ties can share residual mass, with no change in the bound. If the lower vector already sums to one, the maximum coefficient is a valid pivot. The minimum is $-\operatorname{hi}_P(-w)$. Bounds valid coordinatewise for a possibly varying coefficient vector remain valid after applying these extrema to its respective endpoints, since the probability weights are nonnegative.

For $\boldsymbol{v}\in\mathcal D_r$ define $K_{ij}(\boldsymbol{v})=1\{i<j\}+(1/2-v_i)1\{i=j\}$. For each fixed $(\boldsymbol{p},\boldsymbol{q})$ let $\boldsymbol{d}=\boldsymbol{d}^*(\boldsymbol{p}\odot \boldsymbol{q};r)$ and $\mathbf{K}=\mathbf{K}(\boldsymbol{d})$. Put
\[
 R=Kq,\qquad S=\mathbf{K}^{\mathsf T}\boldsymbol{p},\qquad
 L=\boldsymbol{p}^{\mathsf T}Kq=\boldsymbol{p}^{\mathsf T}R=\boldsymbol{q}^{\mathsf T}S
   =\min_{\boldsymbol{v}\in\mathcal D_r}\boldsymbol{p}^{\mathsf T}\mathbf{K}(\boldsymbol{v})\boldsymbol{q} .
\]
Here $L$ is the lower identification endpoint. Values of $R,S$ on zero-probability bins are given by the same formulas. The allocation envelope implies $W^-=\mathbf{K}(d^U)\le \mathbf{K}\le W^+=\mathbf{K}(d^L)$. Thus bounds for the $i$th row mean are $\operatorname{lo}_Q(W^-_{i,\cdot})$ and $\operatorname{hi}_Q(W^+_{i,\cdot})$, denoted by $R_i^-,R_i^+$. Bounds $S_j^-,S_j^+$ use the $j$th column and $P$. The corresponding two lower bounds for $L$ are $\operatorname{lo}_P(R^-)$ and $\operatorname{lo}_Q(S^-)$, and the two upper bounds are $\operatorname{hi}_P(R^+)$ and $\operatorname{hi}_Q(S^+)$.

For a generic box, the sharp cutwise CDF bounds are
\[
 F^-(k)=\max\left(\sum_{j\le k}\ell_j,1-\sum_{j>k}u_j\right),\qquad
 F^+(k)=\min\left(\sum_{j\le k}u_j,1-\sum_{j>k}\ell_j\right).
\]
The two sums constrain the prefix mass from each side. Any value in this range can be realized using the capacity condition separately in the prefix and suffix. These bounds are nondecreasing with endpoints zero and one. Write $\boldsymbol{p}^-,\boldsymbol{p}^+,\boldsymbol{q}^-,\boldsymbol{q}^+$ for distributions with the associated lower and upper marginal CDFs. For fixed $\boldsymbol{v}$, the kernel decreases in $i$ and increases in $j$: the intermediate diagonal value lies in $[0,1/2]$. Couple the three versions of each marginal by its common uniform quantile, independently across groups. A lower CDF gives a larger bin. This yields
\[
 (\boldsymbol{p}^-)^{\mathsf T}\mathbf{K}(\boldsymbol{v})\boldsymbol{q}^+\le \boldsymbol{p}^{\mathsf T}\mathbf{K}(\boldsymbol{v})\boldsymbol{q}
 \le (\boldsymbol{p}^+)^{\mathsf T}\mathbf{K}(\boldsymbol{v})\boldsymbol{q}^-.
\]
Taking the minimum on the common $\mathcal D_r$ preserves the inequalities. The implementation uses the maximum of the two linear lower bounds, the lower stochastic-order corner, and zero, denoted by $L^-$, and the minimum of the two upper bounds, the upper corner, and one, denoted by $L^+$. The same coupling argument applies to any outward rounded CDFs that retain their order and valid endpoints. For a corner with CDF $F_{\boldsymbol{q}}$ its value is evaluated using $\sum_i p_i[1-F_{\boldsymbol{q}}(i)+q_i(1/2-d_i)]$.

\subsection{Simplex boxes and projection-moment envelopes}

For a fixed reference $(\boldsymbol{p}_c,\boldsymbol{q}_c,d_c)$ define $\mathbf{K}_c=\mathbf{K}(d_c)$, $R_c=\mathbf{K}_cq_c$, $S_c=\mathbf{K}_c^{\mathsf T}\boldsymbol{p}_c$ and $L_c=\boldsymbol{p}_c^{\mathsf T}\mathbf{K}_cq_c$. Denote changes from the reference by $\Delta$, and let $e=d_c-d$. Then $\mathbf{K}=\mathbf{K}_c+\operatorname{diag}(e)$, $\Delta R=\mathbf{K}_c\Delta \boldsymbol{q}+\boldsymbol{q}\odot e$, and $\Delta S=\mathbf{K}_c^{\mathsf T}\Delta \boldsymbol{p}+\boldsymbol{p}\odot e$. The reference need only be fixed in the algebra. In particular, all centering below uses the true unit masses.

Put $F_c=R_c-L_c\mathbf{1}$. Since $\boldsymbol{p}^{\mathsf T}(R-L_c\mathbf{1})=\Delta L$,
\[
\begin{aligned}
 n_0 v_0&=\sum_i p_i(F_{c,i}+\Delta R_i)^2-(\Delta L)^2\\
        &=T_1+T_2+T_3+T_4+T_5-(\Delta L)^2,\\
 T_1&=\sum_i p_iF_{c,i}^2,&
 T_2&=2\{\mathbf{K}_c^{\mathsf T}(\boldsymbol{p}_c\odot F_c)\}^{\mathsf T}\Delta \boldsymbol{q},\\
 T_3&=2\sum_i\Delta p_i F_{c,i}(\mathbf{K}_c\Delta \boldsymbol{q})_i,&
 T_4&=2\sum_i c_i F_{c,i}e_i,\qquad
 T_5=\sum_i p_i(\Delta R_i)^2 .
\end{aligned}
\]
This follows by expanding the square, then splitting its cross term using $\Delta R=\mathbf{K}_c\Delta \boldsymbol{q}+\boldsymbol{q}\odot e$ and $\boldsymbol{p}=\boldsymbol{p}_c+\Delta \boldsymbol{p}$. For each coordinate let $D_{\boldsymbol{p},i}$ bound $|\Delta p_i|$ via its two box endpoints. Let $H_{Q,i}$ be the absolute bound on $(\mathbf{K}_c\Delta \boldsymbol{q})_i$ obtained by evaluating the corresponding LP over $Q$ and subtracting $R_{c,i}$. Define $H_{P,j}$ using the columns of $\mathbf{K}_c$ and $P$.

For $T_1$ use the LP on $P$ with squared reference coefficients. For $T_2$, use both extrema over $Q$ of the linear coefficients $2\mathbf{K}_c^{\mathsf T}(\boldsymbol{p}_c\odot F_c)$ and subtract the reference inner product. For $T_3$ use the two-sided absolute bound $A_3=2\sum_i D_{\boldsymbol{p},i}|F_{c,i}|H_{Q,i}$. For $T_4$ take minima and maxima of the four corner products of $c_i$ and $e_i$ times $F_{c,i}$, then sum and scale by $2$. For $T_5$ use zero below and above use the LP over $P$ with coefficients $\max\{(R_i^--R_{c,i})^2,(R_i^+-R_{c,i})^2\}$. Finally, the centering term is in $[-A_L,0]$ with $A_L=\max\{(L^--L_c)^2,(L^+-L_c)^2\}$. If $t_k^\pm$ denote the respective term bounds, then
\[
 \max\{0,(t_1^-+t_2^- - A_3+t_4^--A_L)/n_0\}
 \le v_0\le
 \max\{0,(t_1^++t_2^++ A_3+t_4^++t_5^+)/n_0\}.
\]
For group 1 apply the same identity to the transposed kernel, changing $(\boldsymbol{p},\boldsymbol{q},F_c,\Delta R,n_0)$ to $(\boldsymbol{q},\boldsymbol{p},S_c-L_c\mathbf{1},\Delta S,n_1)$. Sum the two lower bounds and the two upper bounds, respectively, to enclose $V$.

Use $f_i\in[R_i^--L^+,R_i^+-L^-]$ and $g_j\in[S_j^--L^+,S_j^+-L^-]$, denoted by $[f_i^-,f_i^+]$ and $[g_j^-,g_j^+]$. Cubing preserves order for signed arguments. Apply the lower LP to $(f^-)^{\odot3}$ and $(g^-)^{\odot3}$ and the upper LP to the upper endpoints, dividing the group totals by $n_b^2$. For $m_4$, take the largest endpoint absolute value in each coordinate to the fourth power and use the upper probability LP for each margin, dividing by $n_b^3$. Taking the maximum over those endpoint absolute values divided by the corresponding $n_b$ bounds $b$. Including a zero-probability bin in this maximum only increases the upper bound.

Let $D_{c,ij}=\mathbf{K}_{c,ij}-R_{c,i}-S_{c,j}+L_c$. Expanding the mean gives
\[
 \Delta L=R_c^{\mathsf T}\Delta \boldsymbol{p}+S_c^{\mathsf T}\Delta \boldsymbol{q}+
          \Delta \boldsymbol{p}^{\mathsf T} \mathbf{K}_c\Delta \boldsymbol{q}+\sum_k c_ke_k .
\]
Subtracting the row and column changes and adding this mean change gives
\[
\begin{aligned}
 D_{ij}-D_{c,ij}={}&(S_c-\mathbf{K}_{c,i,\cdot}^{\mathsf T})^{\mathsf T}\Delta \boldsymbol{q}
                  +(R_c-\mathbf{K}_{c,\cdot,j})^{\mathsf T}\Delta \boldsymbol{p}\\
 &+\Delta \boldsymbol{p}^{\mathsf T}\mathbf{K}_c\Delta \boldsymbol{q}+1\{i=j\}e_i
                    -q_i e_i-p_j e_j+\sum_k c_ke_k .
\end{aligned}
\]
For the first two terms, use both LP extrema and subtract their reference inner products. For the bilinear term use the absolute bound $\sum_i D_{\boldsymbol{p},i}H_{Q,i}$. The remaining products use the four corners of their component intervals, with subtraction exchanging their two endpoints. Combining these bounds with $D_{c,ij}$ gives a two-sided interval for $D_{ij}$. Let $M_{ij}$ be the maximum of squared interval endpoints. First optimize over $Q$ row by row, and then over $P$. Pointwise,
\[
 E D_{XY}^2\le \sum_i p_i\,\operatorname{hi}_Q(M_{i,\cdot})
       \le \operatorname{hi}_P((\operatorname{hi}_Q(M_{i,\cdot}))_{i=1}^J).
\]
Each row bound holds for every $\boldsymbol{q}$, so they can be summed with nonnegative $\boldsymbol{p}$ even if their individual maximizing distributions differ.

The terms in $W$ have mean zero. Two disjoint pairs are independent. If they share exactly one index, condition on the shared observation; the other observations are independent and the conditional means of $D$ vanish. Thus only the identical pairs contribute to the variance, and $\operatorname{Var}(W)=ED_{XY}^2/(n_0n_1)$. Dividing the nested LP bound by $n_0n_1$ gives the claimed bound. The same conditional mean argument shows $\operatorname{Cov}(W,T)=0$. All arguments permit unequal sample sizes, structural zero bins, a single bin, no overlap, zero budget or saturation.

\FloatBarrier
\section{Fourier and Bennett tail bounds}\label{app:tail-bounds}

This section derives the inclusive projection-tail bounds used by the fixed-pair test. The variance, cumulant, and summand envelopes are those constructed over the probability boxes in Appendix~\ref{app:probability-boxes}. Section~\ref{sec:inference} gives the resulting tail rule.

\subsection{Prawitz smoothing and the signed correction}

Let $S=\sum_\ell Z_\ell$ be the independent mean-zero projection component of the fixed-witness Hoeffding decomposition in Appendix~\ref{app:proofs}. Use $V=\sum_\ell E Z_\ell^2$, $\kappa=\sum_\ell E Z_\ell^3$ and $m_4=\sum_\ell E Z_\ell^4$. In particular $m_4$ refers to the increments; $E S^4$ has the additional cross terms $6\sum_{\ell<k}E Z_\ell^2 E Z_k^2$. Let $\varphi(t)=E e^{itS}$. The notation $V_\pm,\kappa_\pm,m_{4,+},v_{b,+},b_+$ denotes the box envelopes of Appendix~\ref{app:probability-boxes}, where a group-$b$ increment has variance $v_b/n_b$.

Use the convention $\widehat f(u)=\int f(y)e^{-2\pi i u y}dy$. Let $K(y)=(\sin(\pi y)/(\pi y))^2$ with value one at zero. The self-convolution of the unit-length interval indicator gives $\widehat K=(1-|u|)_+$ and $\int K=1$. Its periodization is identically one, by the Fourier coefficients $\widehat K(k)$ and uniform convergence on one period. Define
\[
 H(y)=\sum_{n\ge1}\{K(y-n)-K(y+n)\}+2yK(y),\qquad
 M(y)=(1+H(y)+K(y))/2.
\]
For $y>0$ the identity $1-H(y)=K(y)[1+2y^2\sum_{n\ge1}(y+n)^{-2}-2y]$ follows by subtracting $H$ from the periodization. The decreasing-function sum lies between $1/(y+1)$ and $1/y$; the bracket belongs to $[-1,1]$. Oddness of $H$ handles the other half-axis. At zero $M(0)=1$. Consequently $\Delta(y)=M(y)-1\{y\ge0\}$ lies in $[0,K(y)]$. The odd part is integrable and cancels, giving $\int\Delta=1/2$.

Put $J(u)=(1-u)\pi u\cot(\pi u)+u$ for $0<u<1$. Then
\[
 M(y)=\tfrac12+\int_0^1
 \left[(1-u)\cos(2\pi uy)+\frac{J(u)}{\pi u}\sin(2\pi uy)\right]du.
\]
Indeed insert Abel factors $\rho^n$ into the translates; their Fourier multipliers have limiting odd part $-i\cot(\pi u)$, bounded in modulus by $|\cot(\pi u)|$. After pairing positive and negative frequencies the sine cancels the origin singularity, and $(1-u)$ handles the other endpoint, justifying dominated convergence. The transform of $2yK(y)$ is $(i/\pi)\widehat K'(u)=-i\operatorname{sgn}(u)/\pi$ on the band. Thus the formula includes the exact real and imaginary coefficients. For finite first moments the paired formula can be integrated against the distribution. With $0<u\le1/2$, $0\le\pi u\cot(\pi u)\le1$. For $u>1/2$, $J(u)=u[1-\pi(1-u)\cot(\pi(1-u))]\in[0,u]$. Hence $0\le J\le1$.

For $T>0$, $P(S\ge x)\le E M(T(S-x)/(2\pi))$, including all threshold atoms. Compare the latter linear functional with the transform $\psi(t)=e^{-Vt^2/2}(1+\kappa(it)^3/6)$ of the signed density $f_V-\kappa f_V'''/6$, with $f_V$ the $N(0,V)$ density. The sign follows by integration by parts with the positive characteristic exponent. Changing variables $t=Tu$ and bounding the two coefficients gives a discrepancy at most
\[
 \int_0^T \left(\frac1{\pi t}+\frac1T\right)|\varphi(t)-\psi(t)|\,dt .
\]
Only the probability law uses the pointwise majorant inequality. The signed density enters by equality and an explicit discrepancy bound.

For $x>0$, $V_->0$, set $x_{\mathrm{sh}}=\max(0,x-2\pi R/T)$, $R>0$. On $|T(s-x)/(2\pi)|\le R$, $|s|\ge x_{\mathrm{sh}}$. Since the log-density derivative in $V$ is $(x_{\mathrm{sh}}^2-V)/(2V^2)$, the maximal density there is bounded by $f_{V_*}(x_{\mathrm{sh}})$ with $V_*=\mathrm{clip}(x_{\mathrm{sh}}^2,[V_-,V_+])$. Multiply this by the smoothing mass at most $(2\pi/T)/2$. Outside the window use $\Delta(y)\le1/(\pi^2 y^2)$ and the maximum global density $1/\sqrt{2\pi V_-}$. Adding the monotone normal tail yields
\[
 A_1=\overline\Phi(x/\sqrt{V_+})+\frac{\pi}{T} f_{V_*}(x_{\mathrm{sh}})
         +\frac{4}{\pi T R\sqrt{2\pi V_-}}.
\]
The signed-density step integral is $\kappa f_V''(x)/6=\kappa h(z,V)$ with $h=(z^2-1)\phi(z)/(6V^{3/2})$. For $z=x/\sqrt V$ in its positive interval, the numerator has derivative $z(3-z^2)\phi(z)$. Evaluate both endpoints and the interior critical point $\sqrt3$ when included. For the lower ratio divide a negative minimum by $6V_-^{3/2}$ and a nonnegative minimum by $6V_+^{3/2}$. For the upper ratio divide a positive maximum by $6V_-^{3/2}$ and a nonpositive maximum by $6V_+^{3/2}$. The largest of four products over this interval and $[\kappa_-,\kappa_+]$ bounds the step term. The remaining smoothing contribution is bounded by $(\pi/T)\max|\kappa_\pm|\,0.56/(6V_-^2)$. The positive critical points of $g(z)=(z^3-3z)\phi(z)$ satisfy $z_\pm^2=3\pm\sqrt6$. By symmetry, it is enough to bound $|g|$ at these two positive points:
\begin{align*}
|g(z_-)|
&=\frac{\sqrt6\,\sqrt{3-\sqrt6}}{\sqrt{2\pi}}e^{-(3-\sqrt6)/2}
 < (2.4495)(0.7421)(0.3990)(0.760) < 0.56,\\
|g(z_+)|
&=\frac{\sqrt6\,\sqrt{3+\sqrt6}}{\sqrt{2\pi}}e^{-(3+\sqrt6)/2}
 < (2.4495)(2.335)(0.3990)(0.075) < 0.56.
\end{align*}
Each decimal factor is an outward upper bound: the square-root bounds follow by squaring, and $1/\sqrt{2\pi}<0.3990$ follows from $\pi>3.1415$. The exponential bounds follow from the positive Taylor lower bounds
\[
\sum_{k=0}^{3}\frac{0.27525^k}{k!}>\frac1{0.760},
\qquad
\sum_{k=0}^{6}\frac{2.7245^k}{k!}>\frac1{0.075},
\]
together with $(3-\sqrt6)/2>0.27525$ and $(3+\sqrt6)/2>2.7245$. Thus $\sup_z |(z^3-3z)\phi(z)|<0.56$, with strict inequalities throughout. Zero and infinity contribute zero. This gives $A_2$ without approximating a signed probability.

\subsection{Low-frequency characteristic-function remainder}

For $w_\ell=Ee^{itZ_\ell}-1$, the exponential Taylor integral gives $|w_\ell|\le t^2 E Z_\ell^2/2$ and the cubic remainder at most $t^4 E Z_\ell^4/24$. For $t\le T_1$ set $z_b=T_1^2v_{b,+}/(2n_b)\le 0.4$ in the frozen formulas. The analytic logarithm of each factor is its series at one; $|\log(1+w)-w|\le |w|^2/[2(1-|w|)]$. Exponentiating the sum of these logarithms gives
\[
 \varphi(t)=\exp\{-Vt^2/2+\kappa(it)^3/6+E(t)\},\quad
 |E(t)|\le c_4 t^4,\quad
 c_4=m_{4,+}/24+\sum_{b=0}^1\frac{v_{b,+}^2}{8n_b(1-z_b)} .
\]
For $T_s\le T_1$, set $a=V_-/2$, $C_4=c_4e^{c_4T_s^4}$ and $c_6=\max|\kappa_\pm|^2/72$. For $t\le T_s$, the difference from $\psi$ has modulus at most $e^{-at^2}(C_4t^4+c_6t^6)$, by $|e^E-1|\le |E|e^{|E|}$ and the second-order integral bound on the imaginary axis for $e^{\kappa(it)^3/6}-(1+\kappa(it)^3/6)$. Extending the positive envelope integral to infinity and evaluating the four gamma integrals gives
\[
 A_{3,\mathrm{lo}}=\frac{1}{\pi}\left(\frac{C_4}{2a^2}+\frac{c_6}{a^3}\right)
 +\frac1T\left(\frac{3\sqrt\pi C_4}{8a^{5/2}}+\frac{15\sqrt\pi c_6}{16a^{7/2}}\right).
\]

\subsection{High-frequency bound and Bennett fallback}

For group 0, the centered shift does not change the modulus. Squaring the modulus of its single-increment characteristic function gives $1-2\sum_{i<j}p_ip_j(1-\cos(t(R_i-R_j)/n_0))$. The square root is bounded by the exponential of minus that sum. Multiply across independent increments and groups. Over any interval $[t_a,t_b]\subseteq[T_s,T]$, use the four products of the time endpoints and the enclosed signed row or column difference, divided by its sample size. The minimum of $1-\cos$ on each angular interval is zero if it contains $2k\pi$, and otherwise is the lesser of its endpoint values. Lower probability bounds and these minima supply a nonnegative exponent $E_{ab}$ such that $|\varphi|\le e^{-E_{ab}}$. Also $|\psi|\le (1+t_b^3\max|\kappa_\pm|/6)e^{-at_a^2}$. Multiply their sum by the exact kernel weight $\log(t_b/t_a)/\pi+(t_b-t_a)/T$ and sum over a complete partition. This is $A_{3,\mathrm{hi}}$ (zero when $T_s=T$), and holds over the whole frequency interval. Together all terms bound the inclusive tail. These are finite-sample inequalities for each parameter and every scalar threshold.

The Prawitz kernel is recorded in Section~2 of \citet{Pinelis2013arxiv}. The original citation is \citet{Prawitz1972}; see also \citet{Vaaler1985}.

For $z\le b_+$ and $\lambda\ge0$, the integral remainder for $e^{\lambda z}$ gives
\[
 e^{\lambda z}-1-\lambda z
 \le (z^2/b_+^2)(e^{\lambda b_+}-1-\lambda b_+).
\]
The following bound uses the argument of \citet{Bennett1962}. Taking expectations, using independence and $1+y\le e^y$, then applying Markov at $\lambda=b_+^{-1}\log(1+b_+x/V_+)$ proves the inclusive upper tail bound $\exp(-V_+ h(b_+x/V_+)/b_+^2)$ for $x>0$ and $V_+,b_+>0$, where $h(u)=(1+u)\log(1+u)-u$. Zero variance or zero bound implies the sum is zero. At $x\le0$ use one. The variance floor and scale cutoff in the smoothing routine only decide whether to use the extra bound. Either Bennett alone or the minimum of two valid bounds controls the same probability.

\FloatBarrier
\section{Numerical certification and stopping rules}\label{app:numerical-cert}

This section documents the arithmetic contract, directed checks, and fixed limits used by Cert-Prawitz-BnB. Appendix~\ref{app:proofs} proves the finite-sample level. Appendix~\ref{app:probability-boxes} supplies the probability-box envelopes, and Appendix~\ref{app:tail-bounds} derives the projection-tail bounds used by that proof.

\subsection{Arithmetic model and certified allocation root}

Binary64 inputs denote the exact numbers they represent. Basic operations use correctly rounded arithmetic, round-to-nearest with ties to even, and gradual underflow; excess precision, contraction, and reassociation are disabled. Transcendental bounds use directed MPFR operations.

The saturation check determines the sign of $A(\boldsymbol c)-r$ by exact summation. At a finite scale, the budget residual is
\begin{equation}\label{eq:S4-budget-residual}
 \left(\sum_{t_j>1}a_j-r\right)
 +\sum_{t_j\le1}\frac{a_jt_j^2}{3}
 -\sum_{t_j>1}\frac{2a_j}{3t_j},
 \qquad t_j=\frac{hc_j}{a_j}.
\end{equation}
Directed mantissa products and binary exponent scaling enclose $t_j$. In \eqref{eq:S4-budget-residual}, lower bounds on the high-branch deficit give an upper residual, and upper bounds give a lower residual. The inverse scale evaluates the high branch directly, while monotonicity permits the two bounds to use different branches. Error-free expansions determine the exact signs of rounded sums, following Sections~2.3--2.4 and 2.8 of \citet{Shewchuk1997}. A nonpositive upper residual certifies a lower root bound, and a nonnegative lower residual certifies an upper root bound. The same checks apply to the initial guess, with 0 and $+\infty$ as outer endpoints.

Exact sums determine simplex feasibility and fill residuals. Directed rounding of the linear-program dual gives an upper bound, and negation gives the corresponding lower bound. Moment enclosures use outward intervals for coefficient products, means, and signed moments. Products use their four corners, squares include zero when their interval crosses zero, cubes use monotonicity, and division by a positive interval uses its reversed reciprocal.

\subsection{Confidence-box endpoints and tail checks}

Each confidence-box endpoint is certified by comparing an upper binomial tail with a lower representation of $\gamma/(4J)$. Computation starts from an exact GMP coefficient and upward MPFR powers. Directed recurrence and a geometric bound enclose the remaining positive masses. If an endpoint cannot be certified, the implementation uses 0 or 1 as appropriate.

A lower bound on $\widehat L$, an upper Cantelli allowance, and downward subtraction give $x_B$ in the direction required by the fixed-witness inequality in Appendix~\ref{app:proofs}. Thus numerical rounding preserves the required lower bound on the projection threshold.

For Bennett's bound, the exponent magnitude uses a lower bound on $h$. When $u\le1/8$, the implementation integrates the even truncation of the alternating series for $\log(1+u)$; otherwise it uses a directed MPFR lower bound for the logarithm. This makes the computed tail an upper bound.

For the normal term, monotonicity of $\operatorname{erfc}$ encloses the Gaussian tail. Interval extrema enclose the signed correction, and directed positive operations enclose the low-frequency coefficients when $z_b<1$.

The quarter-power cutoff and bandwidth are represented by admissible real cutoffs. Every interior grid proposal must increase and remain within the exactly enclosed outer endpoints.

For the high-frequency term, directed bounds for $\pi$ and a conservative integer-period check determine the angular intervals. Upper bounds on endpoint cosines then give lower bounds on angular losses. Lower bounds on losses and exponent magnitudes, together with upper bounds on positive exponentials and log weights, enclose all four tail terms from above. Appendix~\ref{app:tail-bounds} gives their analytic derivation.

A box is removed by the tail rule only when its upper tail bound at $x_B$ is no greater than the lower representation of $\delta_P$. Search proposals can set the processing order. They do not certify a rejection.

\subsection{Search limits and conservative returns}

The queue bisects the widest coordinate of the box with the largest recorded proposal, with ties following the fixed code ordering. Both closed children are retained unless a valid certificate removes the box. Table~\ref{tab:certification-limits} lists the fixed limits and the corresponding conservative returns.

\begin{table}[htbp]
\centering
\caption{Certification conditions, implementation limits, and conservative returns.}\label{tab:certification-limits}
\small
\begin{tabularx}{\textwidth}{@{}>{\raggedright\arraybackslash}p{0.47\textwidth}>{\raggedright\arraybackslash}X@{}}
\toprule
Condition & Implementation response\\
\midrule
Initial nonpositive gap or center stop & Return non-rejection\\
Widest coordinate has width below $10^{-5}$ & Return non-rejection\\
256 splits with unresolved boxes & Return non-rejection\\
Geometric test & Require certified lower $L>\tau+10^{-12}$\\
Prawitz variance floor / bandwidth multiplier & $10^{-12}$ / 300\\
Inadmissible smoothing & Use Bennett\\
Nonpositive or nonfinite native tail threshold & Set the tail bound to 1\\
Clopper--Pearson sample size above 100,000 or unverified direction & Use the corresponding endpoint of $[0,1]$\\
\bottomrule
\end{tabularx}
\end{table}

\FloatBarrier

\end{document}